\documentstyle[12pt]{article}
\begin{document}

\title{A Universal Model \\
of D=4 Spinning Particle}
\author{S.L. Lyakhovich, A.Yu. Segal and A.A. Sharapov{\thanks{%
e-mail: sharapov@phys.tsu.tomsk.su}}}
\date{Department of Theoretical Physics, Tomsk State University,\\
Tomsk 634050, Russia}
\maketitle

\begin{abstract}
A universal model for D=4 spinning particle is constructed with the
configuration space chosen as ${\bf R}^{3,1}\times S^2$, where the sphere
corresponds to the spinning degrees of freedom. The Lagrangian includes all
the possible world--line first order invariants of the manifold. Each
combination of the four constant parameters entering the Lagrangian gives
the model, which describes the proper irreducible Poincar\'e dynamics both
at the classical and quantum levels, and thereby the construction uniformly
embodies the massive, massless and continuous helicity cases depending upon
the special choice of the parameters. For the massive case, the connection
with the Souriau approach to elementary systems is established. Constrained
Hamiltonian formulation is built and Dirac canonical quantization is
performed for the model in the covariant form. An explicit realization is
given for the physical wave functions in terms of smooth tensor fields on $%
{\bf R}^{3,1}\times S^2$. One-parametric family of consistent interactions
with general electromagnetic and gravitational fields is introduced in the
massive case. The spin tensor is shown to satisfy the Frenkel-Nyborg
equation with arbitrary fixed giromagnetic ratio in a limit of weakly
varying electromagnetic field. \vspace{1cm}
\end{abstract}

\section{Introduction}

The longstanding problem of the (higher)spin particle dynamics provokes
today the new interest stimulated by the progress of the string theory which
contains the infinite spectrum of higher spin massive excitations. Although
the considerable number of the particle models has been proposed involving
either even or odd spinning degrees of freedom \cite{1}--\cite{18}, the
universal Lagrangian for an arbitrary spin particle in arbitrary dimension
is still in question.

Besides the pure academic interest, this topic seems to be closely related
to the higher spin interaction problem, as for the sake of the consitency
requirements, all the spins have usually to be involved {\it all together}
each time when the attempt is undertaken to introduce the interaction both
in string and field theoretical approaches \cite{27,28}.

It is commonly accepted to construct the particle configuration space as a
direct product of the Minkowski space ${\bf R}^{3,1}$ to an inner manifold
presenting the spinning degrees of freedom. So the first question is to
choose an appropriate space for spin. As a rule, a linear space is used for
this end for the sake of explicit covariance. On the other hand, for the
reasons of the Poincar\'e irreducibility the model should have a proper
number of physical degrees of freedom, that requires to eliminate the extra
variables from the linear space with the aid of constraints. The structure
of the constraints may be rather complicated and thereby obstructs switching
on the interaction or/and prevents to perform the quantization in the
covariant form.

Recently the massive spinning particle model has been proposed with a
two--sphere $S^2$ as the inner space \cite{ms particle}. Although $S^2$ has
a minimal possible dimension among the homogeneous spaces of the Lorentz
group, it does provide the uniform description for an arbitrary spin
dynamics. It should be noted that the group acts on the space {\it %
nonlinearly}, namely by the holomorphic fractional--linear transformations,
i.e. it coincides with the group of complex automorphisms of $S^2$. The
geometric construction of the Ref. \cite{ms particle} gives an explicitly
invariant Lagrangian whose canonical quantization leads to the physical
state subspace being transformed by an irreducible representation of the
Poincar\'e group. This approach is naturally being generalized to the
massive higher superspin superparticle case \cite{super} as well as to the
spinning particle in the constant curvature spaces \cite{AdS}. However, the
action of the Ref. \cite{ms particle} has not involved a possible geometric
invariant of the configuration space, that breaks down the possibility to
introduce a consistent interaction to a general external fields, although
the free theory has been well-defined in itself.

The present paper contains a systematic study of the most general $d=4$
spinning particle model which employes a two--sphere as a configuration
space of the spinning degrees of freedom. {\it All} the possible ${\cal M}^6=%
{\bf R}^{3,1}\times S^2$ world--line Poincar\'e group invariants without
higher derivatives are found including the new Wess--Zumino--like one
unnoticed in Ref. \cite{ms particle}. This invariant transforms under the
Poincar\'e group action by a total derivative. However, its exact invariance
will be recovered if one reformulates the model considering $S^2$ as the
complex projective space ${\bf C}P^1$, wherein the projective
transformations serve as a gauge symmetry of the model. This enables to
construct the most general reparametrization invariant action functional on $%
{\cal M}^6$ describing the particle of an arbitrary fixed spin(helicity).

The action includes four constant parameters, two of which are identified
with mass and spin, and the model is able to describe both massive and
massless particle depending upon the parameters. The rest parameters are
responsible for some arbitrariness in description of the spinning particle
in the framework of the model. In particular, for the massive case the
theory includes two--parametric family of the physically equivalent
Lagrangians of the free particle, however, there is the only special case in
the family, which admits a consistent interaction with general
electromagnetic and gravitational external fields. In this case the spin
tensor is shown to satisfy the Frenkel--Nyborg equation \cite{FrenkelNyborg}
with an arbitrary giromagnetic ratio in the approximation of a weak
electromagnetic field.

Constrained Hamiltonian formulation of the theory is constructed and the
phase space Casimir functions are shown to conserve identically on the
constraint surface. The constraint structure drastically differs in the
general case from the special one, although they describe the equivalent
massive particle dynamics. Whereas the general family is characterized by
two abelian first class constraints, the special case has two second class
constraints and the only first class one. The latter constraint structure is
stable with respect to the deformation of the model by an interaction, while
the first one is not. It is the fact which explains why the only special
case of the model admits the consistent interaction to an external fields.

The global geometry of the reduced phase space is studied. It is shown that
all the independent gauge invariants (observables) of the model are
exhausted by the Hamiltonian generators of the Poincar\'e group. On the
constraint surface, they obey algebraic identities, being phase space
counterparts of operatorial ones characterizing certain irreducible
representation. In fact, the physical phase space of the model can be
identified with the corresponding coadjoint orbit of the Poincar\'e group.
Particularly, this identification has been explicitly established for the
special massive case, where Hamiltonian reduction leads to the dynamics on $%
{\bf R}^6\times S^2$, which is just the group coadjoint orbit being used to
describe the spinning particle dynamics in the framework of Souriau approach
\cite{Souriaubook}. The prequantization condition for the orbit \cite
{Kirillov, Kostant}, being imposed in this framework, restricts the spin of
the massive particle to be (half)integer. Thus, the special case of the
model can be treated as a Lagrangian counterpart of the Souriau description
of the spinning particles.

The basic technical tool applied to realize the Hilbert space of the model
is the relativistic harmonic analysis on $S^2$ \cite{ms particle}, which
enables to provide the manifest covariance of the consideration. By these
means, the canonical quantization of the model leads to the embedding of the
physical state subspace into the space of some smooth tensor fields on $%
{\cal M}^6$. In the framework of the Dirac canonical quantization procedure
\cite{Dirac}, the physical wave function subspace is extracted by quantum
constraint conditions which are proved to coincide with the equations
determining a proper unitary irreducible representation of the Poincar\'e
group.

The relevant Poincar\'e-invariant inner product is constructed to endow the
physical subspace with the Hilbert space structure for all the cases under
investigation: massive, massless with discrete helicity and massless with
continuous helicity. The standard spin--tensor fields on ${\bf R}^{3,1}$
obeying relativistic wave equations appear as coefficients in the expansion
of the physical wave functions via special ``relativistic harmonics'' being
Poincar\'e-covariant generalization of ordinary spherical harmonics.

The paper is organized as follows. In Section 1 we study the geometry of the
configuration space and derive the model in the Lagrangian framework.
Section 2 is devoted to the Hamiltonian formulation of the model. We study
the constraints' structure and identify the parameters of the Lagrangian
with the particle's mass and spin. Also we consider the global geometry of
the physical phase space for the special massive case and establish its
connection to the Kirillov--Kostant--Souriau approach. This special case
allows a consistent interaction to external electromagnetic and
gravitational fields which is studied in Section 3. Section 4 concerns the
quantization of the model. In Section 5 we make some comments and concluding
remarks.

Our notations and conventions coincide mainly with those adopted in the book
\cite{WessBagger} with the exception that, in contrast to Ref .\cite
{WessBagger}, two--component spinor indices are numbered by 0, 1 ($%
\varepsilon ^{01}=\varepsilon ^{{\dot 0}{\dot 1}}=1$) and the spinning
matrices $\sigma _{ab}$ and ${\tilde \sigma }_{ab}$ are defined with the
additional minus sign .

\section{General model of relativistic particle on ${\cal M}^6$}

Similarly to the Minkowski space ${\bf R}^{3,1} $ , ${\cal M}^6$ is a
homogeneous transformation space for the Poincar\'e group, and therefore $%
{\cal M}^6$ can serve as a configuration space where dynamics of
relativistic particle is developed. The requirement of Poincar\'e invariance
strongly restricts the possible choice of an action functional governing a
particle dynamics in ${\cal M}^6$.

In this Section, we classify all the Poincar\'e invariant action functionals
under the following natural conditions :

\noindent i) The Lagrangian does not contain higher derivatives.

\noindent ii) The action is invariant under reparametrizations of the
particle world line.

\noindent iii) The relativistic particle mass-shell condition $p^2=-m^2$
should arise in the theory.

Let us begin with description of ${\cal M}^6$ geometry. In order to present
the action of the Poincar\'e group on ${\cal M}^6$ in the manifestly
covariant form, it is suitable to treat sphere $S^2$ as complex projective
space ${\bf C}P^1$. This space is spanned by nonzero complex two-vector $%
\;\lambda ^\alpha =(\lambda ^0,\lambda ^1)\;$ subject to the equivalence
relation
\begin{equation}
\lambda ^\alpha \sim \kappa \;\lambda ^\alpha \;\;\;,\;\;\;\forall \;\kappa
\in {\bf C}_{*}={\bf C}\backslash \{0\}
\end{equation}
Then, the relations \addtocounter{equation}{1}
$$
x^{\alpha \dot \alpha }\mapsto x^{\prime }{}^{\alpha \dot \alpha }=x^{\beta
\dot \beta }(N^{-1})_\beta \,^\alpha (\overline{N}\,^{-1})_{\dot \beta
}\,^{\dot \alpha }+f^{\alpha \dot \alpha }\eqno{(\theequation.a)}
$$
$$
\lambda ^\alpha \mapsto \lambda ^{\prime }{}^\alpha =\lambda ^\beta
(N^{-1})_\beta \,^\alpha \eqno{(\theequation.b)}
$$
define the action of the Poincar\'e group on ${\cal M}^6$. Here $x^a=-\frac
12(\sigma ^a)_{\alpha \dot \alpha }x^{\alpha \dot \alpha }$ are the
coordinates on ${\bf R}^{3,1}$, $f^a=-\frac 12(\sigma ^a)_{\alpha \dot
\alpha }f^{\alpha \dot \alpha }$ are the parameters of translations, and the
transformations from the Lorentz group $SO^{\uparrow }(3,1)\sim SL(2,{\bf C})%
\big /\pm 1$ are associated in the standard fashion with complex unimodular $%
2\times 2$ matrices
\begin{equation}
\Vert N^\alpha {}_\beta \Vert =\left(
\begin{array}{lr}
a & b \\
c & d
\end{array}
\right) \;\in SL(2,{\bf C})
\end{equation}
In addition to the action of the connected Lorentz group $SO^{\uparrow
}(3,1) $, the discrete transformations $P_1$ and $P_2$, being associated
with space-time reversions, can also be defined on ${\cal M}^6$ by the
relations \addtocounter{equation}{1}
$$
P_1\;:(x^0,x^1,x^2,x^3)\mapsto (x^0,x^1,-x^2,x^3)\;\;,\;\;\lambda ^\alpha
\mapsto \bar \lambda ^{\dot \alpha }\eqno{(\theequation.a)}
$$
$$
P_2\;:x^a\mapsto \,-\;x^a\;\;,\;\;\lambda ^\alpha \mapsto \lambda ^\alpha
\eqno{(\theequation.b)}
$$
Alternatively, one can use the standard complex structure on $S^2={\bf C}%
\cup \{\infty \}$ covered by two charts ${\bf C}$ and ${\bf C}_{*}\cup
\{\infty \}$, with local complex coordinates $z$ and $w$, respectively,
related by the transition function $w=-1/z$ in the overlap of the charts.
The projective coordinates $\lambda ^\alpha $ on $S^2$ are simply connected
with the local ones \addtocounter{equation}{1}
$$
z=\lambda ^1/\lambda ^0\hspace{5mm}\mbox{if}\hspace{5mm}\lambda ^0\not =0%
\eqno{(\theequation.a)}
$$
and
$$
w=-\lambda ^0/\lambda ^1\hspace{5mm}\mbox{if}\hspace{5mm}\lambda ^1\not =0%
\eqno{(\theequation.b)}
$$
In terms of the local coordinates (5.a) the action of the Lorentz group ${%
SO^{\uparrow }(3,1)}$ on $S^2$ is given by the holomorphic fractional-linear
transformations
\begin{equation}
z\mapsto z^{\prime }=\frac{az-b}{-cz+d}
\end{equation}
where $a,b,c,d$ are taken from (3), and for reversions (4) we have
\begin{equation}
P_1\;:z\mapsto \bar z,\hspace{1cm}P_2\;:z\mapsto z
\end{equation}
Thus we arrive to the nonlinear action of the Poincar\'e group on ${\cal M}%
^6 $. In spite of the nonlinearity, the forthcoming description, being
preformed in the local coordinates, has an explicitly covariant form. This
is achieved by making use of the following two-component objects
\begin{equation}
z^\alpha \equiv z^\alpha =(1,z)\;\;,\;\;\bar z^{\dot \alpha }\equiv
(z^\alpha )^{*}=(1,\bar z)\;\;,\;\;\alpha ,\dot \alpha =0,1
\end{equation}
constructed in terms of the local complex coordinates $z$ on $S^2$ and
connected with the projective coordinates $\lambda ^\alpha $ by the rule $%
z^\alpha =\lambda ^\alpha /\lambda ^0,\,\alpha =0,1$. Then the
fractional--linear transformations (6) can be equivalently rewritten in the
covariant form
\begin{equation}
z^\alpha \mapsto z^{\prime }{}^\alpha =\bigg(\frac{\partial z^{\prime }}{%
\partial z}\bigg) ^{1/2}z^\beta (N^{-1})_\beta \,^\alpha
\end{equation}

\noindent This relation means that $\; z^{\alpha}\;$ transforms as a
left-handed Weyl spinor and, simultaneously, as a spinor field on $S^2$.

Describe the Poincar\'e invariants of the tangent bundle $T({\cal M}^6)$.
There are only three independent invariants of the world line :
\addtocounter{equation}{1}
$$
\dot{x}^2 = \dot{x}_a \dot{x}^a\;\; ,\;\; \zeta =\; \frac{%
|\,\lambda_{\alpha} \dot{\lambda}^{\alpha} \,|} {|\,\dot{x}_{\beta\dot{\beta}%
} \lambda^\beta \bar{\lambda}^{\dot{\beta}}\,|} \eqno{(\theequation.a)}%
$$
$$
\Gamma =\rho \frac{\dot{x}_{\alpha\dot{\alpha}} \dot{\lambda}^\alpha \bar{%
\lambda}^{\dot{\alpha}}} {\dot{x}_{\beta\dot{\beta}} \lambda^\beta \bar{%
\lambda}^{\dot{\beta}} } + \mbox{ complex conjugate} \eqno{(\theequation.b)}%
$$

\noindent Here $\rho$ is an arbitrary complex parameter and dot over
variables stands for the derivative with respect to the evolution parameter $%
\tau$ along ${\cal M}^6$ trajectory $\{ x^a (\tau),\lambda^{\alpha} (\tau)
\} $. It should be mentioned that the first two invariants (10.a) have
already been exploited for the construction of the action functional of the
massive spinning particle proposed earlier \cite{ms particle}. Let us make
two remarks concerning the $\Gamma$-invariant which has been unnoticed in
Ref \cite{ms particle}.

\noindent {\it Remark 1}. Strictly speaking, only first two expressions
(10.a), being the true projective invariants with respect to the equivalence
relation $\lambda^\alpha\sim \kappa\:\lambda^\alpha$, are well-defined on $T(%
{\cal M}^6)$, while the last one $\Gamma$ is an invariant modulo a total
derivative:
\begin{equation}
\Gamma \mapsto \Gamma^{\prime}= \Gamma + \frac{d}{d\tau} (\rho\ln\kappa +%
\bar{\rho}\ln\bar{\kappa})
\end{equation}

\noindent {\it Remark 2}. $\Gamma $ is invariant under reversion $\,P_1\,$
only provided $\rho $ is real. At the same time, for a pure imaginary $\rho $%
, $\Gamma $ remains unchanged under the composition $\,P_1\,$ with the
reflection of the evolution parameter : $\tau \rightarrow \,-\tau $.

Even though $\Gamma$ can not be defined on $T({\cal M}^6)$ unambiguously,
one can use it to construct the Poincar\'e invariant action functional for
which the equivalence relation (1) has to serve as a gauge symmetry.

Using invariants (10.a-b) and taking into account the transformation
property of $\Gamma $, one arrives at the most general form of the
Poincar\'e and reparametrization invariant action functional
\begin{equation}
S=\int d\tau {\cal L}=\int d\tau (\sqrt{-\dot x^2{\cal F}(\zeta )}+\Gamma )
\end{equation}
Here ${\cal F}$ is an arbitrary function of $\zeta $ -invariant. Since the
model of a relativistic particle is considered, the mass-shell condition
\begin{equation}
p^2\equiv \frac{\partial {\cal L}}{\partial \dot x^a}\frac{\partial {\cal L}%
}{\partial \dot x_a}=-m^2c^2
\end{equation}
should be imposed. This leads to the following equation for ${\cal F}$ :
\begin{equation}
-{\cal F}+\zeta \,\frac{d{\cal F}}{d\zeta }+4|\rho |^2\zeta ^2=-m^2c^2
\end{equation}
The general solution to the equation is given by
\begin{equation}
{\cal F}=m^2c^2-4\Delta \zeta -4|\rho |^2\zeta ^2\;\;
\end{equation}
with $\Delta $ being an arbitrary real constant of integration.

So we are finally arriving at the most general admissible action
\begin{eqnarray}
S=\int d \tau && \hspace{-7mm}\sqrt{
-\dot{x}^2 \left( m^2 c^2 -4\Delta\
\frac{|\lambda_{\alpha} \dot{\lambda}^{\alpha} |}
{|\dot{x}_{\beta\dot{\beta}} \lambda^\beta \bar{\lambda}^{\dot{\beta}}|}
-4|\rho|^2
\frac{|\lambda_{\alpha} \dot{\lambda}^{\alpha} |^2}
{(\dot{x}_{\beta\dot{\beta}} \lambda^\beta \bar{\lambda}^{\dot{\beta}})^2}
\right)      }
+  \nonumber \\   &{}&  \\
&& + \int d\tau \left\{
\rho \frac{
\dot{x}_{\alpha\dot{\alpha}} \dot{\lambda}^\alpha \bar{\lambda}^{\dot{\alpha}}}
{\dot{x}_{\beta\dot{\beta}} \lambda^\beta \bar{\lambda}^{\dot{\beta}} }
+
\bar{\rho} \frac{
\dot{x}_{\alpha\dot{\alpha}} \lambda^\alpha \dot{\bar{\lambda}}^{\dot{\alpha}}}
 {\dot{x}_{\beta\dot{\beta}} \lambda^\beta \bar{\lambda}^{\dot{\beta}} }
\right\} \nonumber
\end{eqnarray}

\noindent
which depends on three arbitrary parameters : $m^2$ , $\Delta$, and $\rho$.

The action (16) is seen to be manifestly Poincar\'e invariant and possesses
local symmetries. First of all, there are the world-line reparametrizations
\begin{equation}
\delta_{\epsilon} x^a =\dot{x}^a \epsilon\;\;,\;\; \delta_{\epsilon}
\lambda^{\alpha} = \dot{\lambda}^{\alpha} \epsilon
\end{equation}
and local $\lambda^{\alpha}$-rescalings
\begin{equation}
\delta_{\kappa} \lambda^{\alpha} = \kappa\;\lambda^{\alpha}\;\;,\;\;
\delta_{\kappa} x^a =0
\end{equation}
Then, for $\Delta\ne 0$ there exists one more gauge transformation of the
form
\begin{equation}
\delta_{\mu} x^a =\frac{\partial {\cal L}}{\partial \dot{x}_a} \mu\;\;, \;\;
\delta_{\mu} \lambda^{\alpha} =0
\end{equation}
which directly follows from the mass-shell condition. Whereas for $%
\Delta=m(\rho-\bar{\rho})=0$, instead of (19), one finds two gauge
transformations which can be presented in the manner
$$
\addtocounter{equation}{1} \delta_\nu \lambda^\alpha =\frac{1}{2}\bar{\lambda%
}_{\dot{\alpha}} \dot{x}^ {\dot{\alpha} \alpha} \sqrt{-\frac{{\cal F}(\zeta)%
}{\dot{x}^2}} \nu\;\; , \;\; \delta_\nu \bar{\lambda}^{\dot{\alpha}} =0%
$$
$$
\eqno{(\theequation)}%
$$
$$
\delta_\nu x^a =\left(\frac{(\sigma^a)_{\alpha \dot{\alpha}} \lambda^\alpha
\bar{\lambda}^{\dot{\alpha}}} {\lambda_{\beta} \dot{\lambda}^{\beta}}
|\rho|^2 \zeta^2 \sqrt{-\frac{\dot{x}^2}{{\cal F}(\zeta)}} - \rho \frac {%
\dot{x}_b (\tilde{\sigma}^{ba})_{\dot{\alpha} \dot{\beta}} \bar{\lambda}^{%
\dot{\alpha}} \bar{\lambda}^{\dot{\beta}}} {\dot{x}_{\gamma \dot{\gamma}}
\lambda^\gamma \bar{\lambda}^{\dot{\gamma}}}\right)\nu
$$
where $\nu(\tau)$ is a complex parameter.

Locally, the action (16) can be readily rewritten in terms of
projective-invariant coordinates (5.a) : $\lambda ^\alpha $ is represented
as \addtocounter{equation}{1}
$$
\lambda ^\alpha =\kappa z^\alpha \eqno{(\theequation.a)}
$$
then its substitution to (16) gives
$$
S^U=\int d\tau \sqrt{-\dot x^2\left( m^2c^2-4\Delta \frac{|\dot z|}{|(\dot
x,\xi )|}-4|\rho |^2\frac{|\dot z|^2}{(\dot x,\xi )^2}\right) }+
$$
$$
{}
$$
$$
+\int d\tau \left\{ \rho \dot z\partial _z\ln (\dot x,\xi )+\bar \rho \bar
z\partial _{\bar z}\ln (\dot x,\xi )\right\} +\eqno{(\theequation.b)}
$$
$$
{}
$$
$$
+\int d\tau \frac d{d\tau }(\rho \ln \kappa +\bar \rho \ln \bar \kappa )
$$
where
\begin{equation}
\xi _a=(\sigma _a)_{\alpha {\dot \alpha }}z^\alpha {\bar z}^{\dot \alpha }={%
\bar \xi }_a=(1+z{\bar z},z+{\bar z},{\rm i}z-{\rm i}{\bar z},1-z{\bar z}%
),\quad \xi ^a\xi _a=0.
\end{equation}
Here $U$ stands for an open domain in ${\cal M}^6$ parametrized by the
coordinates $(x^a,z)$. In this form the action functional (21.b) is
invariant under the Poincar\'e transformations where the action of the
Lorentz group (6) is completed by the corresponding transformation for $%
\kappa $
\begin{equation}
\kappa \rightarrow \kappa ^{\prime }=\bigg(\frac{\partial z^{\prime }}{%
\partial z}\bigg)^{-1/2}\kappa
\end{equation}

In what follows, the last term in (21.b) containing $\kappa $ is being
omitted as it represents the total derivative, but in so doing the
Lagrangian becomes invariant under the Lorentz transformation up to a total
derivative only. The remaining local symmetries of the action (21) can be
read off from the Eqs. (17),(19),(20) by making formal replacement $\lambda
^\alpha \rightarrow z^\alpha $.

It is interesting to mention that the local form of the invariants
$$
\zeta ^2=\frac{|\dot z|^2}{|(\dot x,\xi )|^2}\qquad \;,\;\qquad \Gamma =\rho
\dot z\partial _z\ln (\dot x,\xi )+\mbox{c.c.}
$$
admits a neat geometrical interpretation connected with the K\"ahler
manifold structure on $S^2$. Defining in a domain $U$ the ${\bf R}^{1,3}$
velocity-dependent K\"ahler potential
$$
\phi =\ln (\dot x,\xi )
$$
one finds the metric
$$
2g_{z\bar z}dzd\bar z=4dzd\bar z\partial _z\partial _{\bar z}\phi \qquad
\;,\qquad \;-2\dot x^2\zeta ^2={g_{z\bar z}}\frac{dz}{d\tau }\frac{d\bar z}{%
d\tau }
$$
and the 1-form potential
$$
\theta =\rho \partial _z\ln (\dot x,\xi )dz+\mbox{c.c.}\qquad \;,\qquad
\;\Gamma {d\tau }=\theta
$$
for the symplectic structure
$$
\omega =-2i\,\Im m\rho \,\partial _z\partial _{\bar z}\phi \,dz\wedge d\bar
z.
$$
In the rest reference system for $\dot x^a=(\dot x^0,0,0,0)$ one comes to
the standard metric on $S^2$
$$
2g_{z\bar z}dzd\bar z=4\frac{dzd\bar z}{(1+z\bar z)^2}.
$$
This geometrical background of the Lagrangian construction (21.b) will be
some more discussed below when the Hamiltonian formalism for the model is
being analyzed.

In the conclusion of this section it is worth to note that the parameters $%
\Delta $ and $\rho $ are dimensional, and they can not be made dimensionless
with the help of the constants $m$ and $c$ only. Accounting, however, the
Planck constant, $\Delta $ and $\rho $ may be represented as
\begin{equation}
\Delta =\hbar mc\delta \;\;,\;\;\rho =\hbar r
\end{equation}
where $\delta $ and $r$ are dimensionless numbers. In what follows, the
parameters $\delta $ and $r$ will be connected with spin or helicity of the
particle. The appearance of $\hbar $ in the classical action seems to be a
quite natural phenomenon from the common standpoint that spin should
manifest itself as a quantum effect disappearing in the classical limit $%
\hbar \to 0$. This fact is not being emphasised in the further
consideration, where we put $\hbar =c=1$.

\section{ Hamiltonian formalism}

Now, we are going to construct a constrained Hamiltonian formulation for the
model.

Beginning with the general Lagrangian (16), introduce conjugated momenta for
the coordinates $x^a,\lambda^{\alpha}$ and $\bar{\lambda}^ {\dot{\alpha}}$

\begin{equation}
p^a= \frac{\partial {\cal L}}{\partial \dot{x}^a} \;,\; \pi_{\alpha} =\frac{%
\partial {\cal L}}{\partial \dot{\lambda}\,^{\alpha}} \;,\; \bar{\pi}_{\dot{%
\alpha}} =\frac{\partial {\cal L}}{\partial \dot{\bar{\lambda}}\,^{\dot{%
\alpha}}} \;,\;
\end{equation}
subject to the canonical Poisson brackets relations
\begin{equation}
\{x^a ,p_b \} = \delta^a_b \;\;,\;\; \{\lambda^{\alpha} ,\pi_{\beta} \}
=\delta^{\alpha}_{\beta}\;\;,\;\; \{ \bar{\lambda}^{\dot{\alpha}} ,\bar{\pi}%
_{\dot{\beta}} \} =\delta^{\dot{\alpha}}_{\dot{\beta}}
\end{equation}

Canonical Hamiltonian of the model vanishes identically by virtue of the
repara\-metrization invariance. Therefore, the model is completely
characterized by the Hamiltonian constraints. It turns out that number and
structure of the constraints essentially depends on the values of the model
parameters. Moreover, as it has already been mentioned, the arbitrariness in
the choice of parameters' values proves to cover all Poincar\'e-irreducible
particle dynamics: massive, massless (including continuous helicity case),
and tachion\footnote{%
In the present paper, we leave the tachion case aside.}.

First of all, the mass-shell constraint
\begin{equation}
T_1=p^2+m^2\approx 0
\end{equation}
always appears for the action (16); then, the constraints corresponding to
the equivalence relation (1) arise
\begin{equation}
d=\lambda ^\alpha \pi _\alpha -\rho \approx 0\;\;,\;\;\bar d=\bar \lambda
^{\dot \alpha }\bar \pi _{\dot \alpha }-\bar \rho \approx 0\;\;,
\end{equation}
and at last in the case of $\;\Delta \neq 0$ there is one more constraint of
the form \addtocounter{equation}{1}
$$
T_2=\theta \bar \theta -\Delta ^2\approx 0\eqno{(\theequation.a)}
$$
where
$$
\theta =\bar \lambda _{\dot \alpha }p^{\dot \alpha \alpha }\pi _\alpha
\;\;,\;\;\bar \theta =\bar \pi _{\dot \alpha }p^{\dot \alpha \alpha }\lambda
_\alpha \eqno{(\theequation.b)}
$$
The complete set of the constraints : $T_1$, $T_2$ , $d$ and $\bar d$
generates an Abelian algebra with respect to the Poisson brackets. Hence,
for nonzero $\Delta $ we get four constraints of the first class.

In the special case of $\Delta =0$ the situation changes drastically.
Instead of one constraint $T_2$ there emerge two additional constraints
\begin{equation}
\theta \approx 0\;\;,\;\;\bar \theta \approx 0
\end{equation}
They possess vanishing Poisson brackets with $T_1$ , $d$ and $\bar d$, and
nonzero bracket among themselves
\begin{equation}
\{\theta ,\bar \theta \}=p^2(\lambda ^\alpha \pi _\alpha -\bar \lambda
^{\dot \alpha }\bar \pi _{\dot \alpha })\approx -m^2(\rho -\bar \rho )
\end{equation}
This implies that, always, excepting $m^2(\rho -\bar \rho )=0$, the
constraints $\theta $ and $\bar \theta $ are of the second-class and reduce
to the first-class ones whenever combination $m^2(\rho -\bar \rho )$
vanishes.

Notice that the constraints $T_i , i=1,2$ have a simple group theoretical
interpretation which may, in particular, assign the clear physical sense to
the parameters entering the Lagrangian (16). Indeed, the Poincar\'e group
acts on the phase space by canonical transformations, and the corresponding
Hamiltonian generators, being constructed as a N$\ddot{\mbox{o}}$ther
charges, look like \addtocounter{equation}{1}
$$
{\cal P}_a =p_a \;\;,\;\;{\cal J}_{ab} =x_a p_b -x_b p_a + {\cal M}_{ab}
\eqno{(\theequation.a)}%
$$
$$
{\cal M}_{ab} = (\sigma_{ab})_{\beta} {\!}^{\alpha} \lambda^{\beta}
\pi_{\alpha} - (\tilde{\sigma}_{ab})^{\dot{\beta}} {\!}_{\dot{\alpha}} \bar{%
\lambda} ^{\dot{\alpha}}\bar{\pi}_{\dot{\beta}} \eqno{(\theequation.b)}%
$$
Then the Casimir functions for Poincar\'e generators (32) read
\addtocounter{equation}{1}
$$
C_1 ={\cal P}^a {\cal P}_a \approx -m^2 \eqno{(\theequation.a)}%
$$
$$
C_2 = {\cal W}_a {\cal W}^a =\theta\bar{\theta} + \frac{1}{4}p^2 (
\lambda^{\alpha} \pi_{\alpha}-\bar{\lambda}^{\dot{\alpha}} \bar{\pi}_{\dot{%
\alpha}})^2 \approx \Delta^2 + m^2 (\Im m \rho)^2 \eqno{(\theequation.b)}%
$$
where ${\cal W}^a =\frac{1}{2}\varepsilon^{abcd} {\cal P}_b {\cal J}_{cd}$
is the classical Pauli-Lubanski vector. Thus, Eqs. (27-30) imply that $C_1$
and $C_2$ are identically conserved on the total constrained surface. Eq.
(33.b) also shows that, for nonzero mass, the value
\begin{equation}
\frac{\Delta ^2}{m^2} +( \Im m \rho)^2
\end{equation}
has the meaning of the squared spin. At the same time, in the case of $%
m=\Delta =0$ it is a simple exercise to check that the set of constraints
(28),(30) is equivalent to the condition
\begin{equation}
{\cal W}^a - (\Im m \rho) {\cal P}^a \approx 0
\end{equation}
and, reversely, Eq. (35) leads to the mentioned constraints.

From the above consideration it follows that there are four essentially
different cases depending on the values of the parameters $\Delta$ , $m$ and
$\rho$. Each case is identified with some Poincar\'e-irreducible dynamics.
They can be summarized as follows~:

\vspace{4mm} \noindent i) $\Delta \neq 0 ,m \neq 0 \;$

\noindent In this case the model is described by the four first-class
constraints and thereby the number of the physical degrees of freedom comes
out to $8-4= 4 = 3\;$(position) + $1\;$(spin). So, one may see that this
case corresponds to the massive spinning particle with squared spin
magnitude $s^2$ given by (34). For the case $\Im m\rho =0$ the model
coincides with the one studied earlier, and it was demonstrated that the
canonical quantization of the theory leads to the equations for the physical
wave functions which prove to be equivalent to the relativistic wave
equations for massive spin-$s$ field \cite{ms particle}.

\vspace{4mm} \noindent ii) $\Delta = 0$ , $m (\rho -\bar{\rho}) \neq 0 \;$

\noindent Despite the change of the constraints' structure (there are three
first and two second-class constraints), the number of physical degrees of
freedom is the same as in the previous case. The model still describes
massive spinning particle whose spin value, in accordance with (34), is
equal to $|\Im m\rho |$. As it will be shown in the Sec. 5, the canonical
quantization procedure, as applied to this case, leads to the irreducible
mass$-m$ , spin$-s$ representation of the Poincar\'e group. This case is
been considered in the paper with most detailes since {\it it is the only
case which allows to introduce a consistent interaction both with
electromagnetic and gravitational fields } (see Sec.4).

\vspace{4mm} \noindent iii) $\Delta = 0 $ , $m = 0 $

\noindent In this case, there are five first-class constraints. Hence, the
number of the physical degrees of freedom equals three. From the Rel. (35)
it follows that the model describes the relativistic massless particle with
helicity $\Im m\rho $. In the Sec. 5 we show that the canonical quantization
of this model leads to the relativistic wave equations for massless,
helicity $\Im m\rho $ field.

Also, it is evident from (16) that, provided $\rho$ is pure imaginary, the
space reflection $P_1$ reverses the helicity's sign, as it must happen in a
massless particle case.

\vspace{4mm} \noindent i$\vee$) $\Delta \neq 0 \,,\,m = 0 $

\noindent This situation is special: the mass is zero while the spin is not
zero. The parameter $\Delta $ can not longer be made dimensionless (see
Rel.(24)), and so new dimensional parameter has to be introduced. The model
possesses four physical degrees of freedom. As it becomes clear when the
theory is being quantized (see Sec.5), this case corresponds to the massless
particle with a continuous helicity.

\vspace{4mm} \noindent In the remained case ($\Delta=\Im m \rho =0$ ) the
model describes a spinless particle.

Let us now discuss the general structure of physical observables in the
theory. Each physical observable $A(x^a,p_b,z,p_z,\bar{z},p_{\bar{z}})$,
being gauge invariant function on the phase space should meet the
requirement
\begin{equation}
\{ A_{phys},\mbox{(first-class constraints)} \} \approx 0
\end{equation}
It turns out that the basis of physical observables is formed by the
generators of Poincar\'e group, i.e. any physical observable of the theory
is a function of the Poincar\'e generators only
\begin{equation}
A_{phys} = A ({\bf {\cal P}}_a\;,\;{\bf {\cal J}}_{ab})
\end{equation}
To prove this assertion consider, first, the cases (i),(ii), and (i$\vee$).
For all these cases the dimension of physical phase space is equal to 8. On
the other hand, the constrained surface is obviously Poincar\'e invariant
that provides all the Poncar\'e generators to be the physical observables.
As a result, the physical phase space can be covariantly parametrized by 10
generators (32) subject to 2 Casimir conditions (33). As for the remained
case (iii), the phase-space, being reduced by the constraints, is
six-dimensional that is agreed well with 10 Poincar\'e generators subject to
4 first class constraints (35).

In the previous section we have described two equivalent Lagrange
formulations for the model: in terms of projective (16) and local (21)
coordinates on the sphere. In this regard it is pertinent to note that the
Hamiltonian formalism for the second case, once constructed, leads to the
essentially similar results to that just considered. The only difference is
that the first-class constraints (28) are satisfied automatically now, so
that the rest constraints are straightforwardly expressed in terms of the
cotangent bundle $T ^* ({\cal M} ^6)$ variables. Let $p_z$ be a canonically
conjugated momentum to the local complex coordinate on the sphere $z$
\begin{equation}
\{z,p_z\}=\{\bar{z},p_{\bar{z}}\}=1
\end{equation}
Then the constraints (27),(29),(30) can be rewritten as
\addtocounter{equation}{1}
$$
T^{\prime}_1\; = T_1 \approx 0 \eqno{(\theequation.a)}%
$$
$$
T^{\prime}_2 =2g^{z\bar z} \nabla_z \nabla_{\bar z} - (\Delta^2 +m^2(\Im m
\rho)^2) \approx 0 \eqno{(\theequation.b)}%
$$
$$
\theta ^{\prime}=(p,\xi) \nabla _z \approx 0 \; , \; \bar\theta
^{\prime}=(p,\xi) \nabla _{\bar z} \approx 0 \eqno{(\theequation.c)}%
$$
where
\begin{equation}
\nabla_z = p_z -\rho \partial_z \ln (p,\xi)\;\;,\;\; \nabla_{\bar{z}} = p_{%
\bar{z}} -\bar{\rho}\partial_{\bar{z}} \ln (p,\xi),
\end{equation}

As before, the constraints $T^{\prime}_1$ and $T^{\prime}_2$ represent the
Casimir functions for the Poincar\'e generators (32.a), wherein the spinning
part of Lorentz transformations are now realized in the manner
\addtocounter{equation}{1}
$$
M_{ab} = (\sigma_{ab})_{\alpha\beta}M^{\alpha\beta} - ({\tilde\sigma}_{ab})_{%
{\dot\alpha}{\dot\beta}} \vspace{-12pt}{\bar M}^{{\dot\alpha}{\dot\beta}},
\eqno{(\theequation.a)}%
$$
$$
{}%
$$
$$
M^{\alpha\beta} = -z^\alpha z^\beta p_z + \frac{\rho}{2} \partial_z
(z^\alpha z^\beta), \qquad {\bar M}^{{\dot\alpha}{\dot\beta}} = - {\bar z}%
^{\dot\alpha} {\bar z}^{\dot\beta} p_{\bar z} + \frac{\bar{\rho}}{2}
\partial_{\bar z}({\bar z}^{\dot\alpha} {\bar z}^{\dot\beta}).
\eqno{(\theequation.b)}%
$$

Let us focus now upon the case (ii), where the model has only one
independent gauge transfomation corresponding to the reparametrization
invariance, that makes possible to introduce the minimal coupling of the
model to external gravitational and electromagnetic fields (see Sec. 4).
Because of the physical importance of this fact, it seems instructive to
consider the structure of reduced phase space corresponding to the free
particle in more detailes.

The constraints (39.c) extract the 9-dimensional surface $E$ in the
cotangent bundle $T^{\ast}( {\cal M}^6)$. Topologically, the manifold $E$ is
equivalent to the ${\bf R}^7 \times S^2$. The restriction of the canonical
symplectic structure $T^{\ast}( {\cal M}^6)$
$$
2(dp_a \wedge dx^a + dp_z \wedge dz + dp_{\bar z} \wedge d{\bar z} ) \, ,
$$
to the constraint surface induces on $E$ the following 2-form
\addtocounter{equation}{1}
$$
\Omega^E = 2 dp_a \wedge dx^a + \eqno{(\theequation)}
$$
$$
+(\Im m \rho) \left( -4im^2 \frac{dz \wedge d \bar{z}}{(p,\xi)^2}+ 2i\frac{{%
p^{\alpha}}_{\dot{\beta}} \bar{z}^{\dot{\alpha}} \bar{z}^{\dot{\beta}}}{%
(p,\xi)^2} dz \wedge dp_{\alpha\dot{\alpha}} -2i\frac{{p_{\beta}}^{\dot{%
\alpha}}z^{\alpha} z^{\beta}}{(p,\xi)^2} d\bar{z} \wedge d p_{\alpha\dot{%
\alpha}}\right)
$$
where $p_a$ obeys the constraint $p^2=-m^2$. Note that the second term in
the expression for $\Omega^E$ may be regarded as a `relativistic'
generalization of the usual symplectic structure on the sphere. Indeed, let
us restrict $\Omega^E$ to a surface $dp_a =0$, then in the rest reference
system where $p_a={\stackrel{\circ}{p}}_a= (m,\vec{0})$, the expression (42)
coincides with the standard symplectic structure on a sphere
\begin{equation}
\left. m^2 \frac{4idz\wedge d\bar z}{(p,\xi)^2}\right| _{p= \stackrel{\circ}{%
p}}=\, \frac{4idz\wedge d\bar z}{(1+z\bar z)^2}
\end{equation}
Hence the sphere $S^2$ appears to be both the configuration and the phase
space for the spin (after Hamiltonian reduction, for the latter).

It is evident that the 2-form $\Omega^E$ is degenerate and the corresponding
foliation $ker\; \Omega^E $ coincides with integral curves of the vector
field
\begin{equation}
\vec{v} =p^a \frac{\partial}{\partial x^a}
\end{equation}
The symplectic reduction procedure is completed through the factorization of
$E$ over the action of the vector field $\vec{v}$. This can be explicitely
done by means of a gauge fixing condition for the constraint $p^2 =-m^2 $,
for example
\begin{equation}
x^0 =t
\end{equation}
where $t$ is an arbitrary parameter. The resulting manifold $\;\Sigma \sim
E\,/ker \Omega\,^E $, being topologically equivalent to ${\bf R}^6 \times
S^2 $, carries a nondegenerate symplectic structure $\;\Omega^{\Sigma}\;$
obtained by the restriction of $\Omega^E$ to the surface (45). So, we
finally conclude that for the case (ii) the physical phase space of the
model is the symplectic manifold $(\Sigma ,\Omega^{\Sigma})$.

Let us now return to the presymplectic manifold $E$. Since $E$ possesses
nontrivial 2-cycles associated with $S^2$, the closed 2-form $\Omega ^E$ is
not exact. This, in particular, implies that its potential can be defined
only locally. In an open domain ${U}\subset E$ having a spherical part
parametrized by the local complex coordinate $z$ the potential $\omega $ can
be chosen as
\begin{equation}
\omega =2(p_adx^a+\rho dz\partial _z\varphi +\bar \rho d\bar z\partial
_{\bar z}\varphi )\;\;,\,\,\;\;\;d\omega =\Omega ^E
\end{equation}
where
\begin{equation}
\varphi =\ln (p,\xi )
\end{equation}
\noindent plays the role of the K\"ahler potential for the `relativistic'
symplectic structure (43) on $S^2$. Using the 1-form $\omega $ one can write
down the Hamiltonian action of the model as
\begin{equation}
S\;^U=\frac 12\int \omega
\end{equation}

Although functional $S\;^{U}$ is defined only for those trajectories which
do not pass through the singular point $z=\infty \not\in {U}$, it leads to a
proper classical dynamics on the whole $E$. The variational principle, as
applied to $S\;^{U}$~, determines the particle's trajectories as the
foliation $ker\; \Omega^E$.
\begin{equation}
\frac{\delta S\;^{U}}{\delta\Gamma^i} =\Omega^E_{ij}\;\dot{\Gamma}^j =0
\end{equation}
where $\Gamma^i=(x^a, p_b , z,\bar z)$. Note that the Eqs. (49) are
well-defined on $E$ as well as $\Omega^E$ is. Taking into account that $%
ker\; \Omega$ is generated by the vector field (44) one can rewrite Eqs.
(49) in the equivalent form
\begin{equation}
\dot{x}^a =\lambda\, p^a \;,\hspace{5mm} \dot{p}^a =0 \;,\hspace{5mm} \dot{z}
=0
\end{equation}
where $\lambda $ is an arbitrary Lagrange multiplier corresponding to the
reparametrization invariance. It follows from (50) that the free massive
particle (ii) keeps rest on the sphere and, in Minkowski space, the
trajectories coincide with time-like geodesic lines\footnote{%
So this case does not exhibit Zitterbewegung phenomenon which appears in the
case (i) \cite{ms particle}.}.

It is pertinent to discuss here the relationship between the above
consideration and the well-known Kirillov-Kostant-Souriau approach \cite
{Kirillov, Souriaubook, Kostant} to classical elementary systems (the
so-called {\it orbit method}). The basic objects in this approach are the
coadjoint orbits ${\cal O}$ of a Lee group ${\cal G}$ for which a system is
`elementary'. The coadjoint orbit is known to be a homogeneous symplectic
manifold, so it may be naturally identified with the physical phase space of
the system. Applied to the Poincar\'e group this method gives the Souriau
classification of free relativistic particles \cite{Souriaubook}. In the
case of mass-$m$ spin-$s$ particle the corresponding orbit ${\cal O}_{m,s}$
exactly coincides with the manifold $(\Sigma ,\Omega ^\Sigma )$ which has
just emerged from the action functional (21.b) in the case (ii) through the
conventional Dirac analysis and the Hamiltonian reduction with the aid of
constraints.

The geometrical quantization \cite{Kostant} of such `an elementary system'
implies that the symplectic structure, being associated to the orbit ${\cal O%
}$, must satisfy the prequantization condition requiring that the integral
of $\Omega $ over an arbitrary 2-cycle is a multiple $2\pi $ :
\begin{equation}
\int\limits_{2-cycle}\Omega =2\pi n\;,\hspace{5mm}n\in {\bf Z}
\end{equation}
Owing to the topological structure of $\Sigma \sim {\bf R}^6\times S^2$, the
only nontrivial 2-cycle corresponds to the sphere in the case under
consideration. This leads to the condition
\begin{equation}
\label{gq}\int\limits_{2-cycle}\Omega ^\Sigma =-m^2\Im m\rho
\,\int\limits_{S^2}\frac{4i\,dz\wedge d\bar z}{(p,\xi )^2}=2\pi
n\;\;,\;\;\;\;n\in {\bf Z}
\end{equation}

\noindent Integrating over sphere and taking into account the constraint $%
p^2=-m^2$ we come to the restriction on an admissible value of spin
\begin{equation}
s=|\Im m \rho| =\frac{n}{2}\;\;\;, n=0,1,2,\dots
\end{equation}

It can be also shown that the quantization rule (53) for the spin parameter $%
s$ provides the expression $\exp iS\;^U$ to be single-valued, being taken on
the trajectories whose projections to $S^2$ are closed curves. The last fact
makes possible to define path-integral correctly in this model along the
lines of Ref.\cite{Faddeev}, where the pure $S^2$ case has particularly been
described. In Sec. 5 we consider the Dirac canonical quantization uniformly
in all the cases of the model in the operator approach, and in so doing the
quantization rule for spin (53) emerges from the analysis of the spectrum of
constraint operators.

\section{ Coupling to external fields}

So far we have discussed the free relativistic spinning particle propagating
in the flat spacetime and have seen that, for each certain choice of the
parameters $m$ , $\Delta $, $\rho $, the model describes a proper
Poincar\'e-irreducible dynamics. Now we proceed to a generalization of the
theory to the case of presence of external gravitational and electromagnetic
fields.

It is well known that there exists a consistency problem in higher spin
field dynamics coupled to gravity and electromagnetism \cite{higherspins}.
One may expect to find a similar problem in the context of the particle
mechanics too. We are going to show that this happens in the model (16)
indeed, and there is the case (ii) only, where a consistent coupling can be
introduced to the general external fields for the massive particle.

It is convenient to consider this problem starting with the Hamiltonian
formulation for the free model described by the constraints (27--30). In
what follows ${e_m}^a(x)$~, $\omega _{m\,ab}(x)$ and ${\cal A}_m(x)$ are the
vierbein, spin connection and electromagnetic potential, respectively. Then
the simplest way to introduce an interaction known as a minimal coupling
consists in the replacement of the momentum $p_a$ by its covariant
generalization $\Pi _a:$
\begin{equation}
p_a\mapsto \Pi _a=e^m\,_a(p_m+e{\cal A}_m+\frac 12\omega _{m\,cd}{\cal M}%
^{cd})
\end{equation}
in all constraints (27--30). Here $e^m\,_a$ is the inverse vierbein, $e$ is
electric charge and the Hamiltonian generators of the Lorentz
transformations ${\cal M}^{ab}$ are defined as in Rel.~(32.b).
Geometrically, the last expression assumes that we replace the `flat'
configuration space of the model ${\cal M}^6={\bf R}^{3,1}\times S^2$ by the
Lorentz bundle $M^4\times S^2$ over curved spacetime $M^4$ with a Lorentz
connection $\omega _{m\,ab}$. In so doing the obtained model turns out to be
invariant under the gauge transformations of external fields since they may
be induced by a canonical transformation.

Notice however that, contrary to $p_a$, the extended momenta $\Pi_a$ possess
nonzero Poisson bracket among themselves
\begin{equation}
\{\Pi_a, \Pi_b\} =e{\cal F}_{ba} + {\cal T}^c_{ba} \Pi_c + \frac{1}{2} {\cal %
R}_{bacd} {\cal M}^{cd}
\end{equation}
where ${\cal R}_{abcd} $ , ${\cal T}^c_{ab}$ and ${\cal F}_{ab}$ are,
respectively, Riemann curvature, torsion and electromagnetic strength
tensors. As a result, the Poisson brackets of the abelian first-class
constraints $T_1$ , $T_2$ corresponding to the cases (i),(i$\vee$) and $%
\theta$ , $\bar{\theta}$ of the case (iii) become proportional to the r.h.s.
of Rel. (55) after substitution (54). This indicates that the coupling,
being introduced in such a way, is inconsistent with the gauge invariance
underlying the free model, and so the resulting theory proves to be
contradictory. What is more, one may show that it is also impossible to
preserve the involution relation of the first-class constraints even by
adding nonminimal terms to them.

The only occurrence, when the construction (54) works well, is the case
(ii). This fact lies mainly in the exceptional algebraic structure of the
constraints in the case. Let us explain this in more details. First, all the
constraints (27~--~30) transform homogeneously under the dilatations
generated by $d$ and $\bar d$. Therefore, both for the free and coupled
models, the constraints $d$ and $\bar d$ are of the first class and so may
be omitted from the subsequent analysis. (Moreover, passing to the local
parametrization these constraints can be explicitly resolved). Then consider
the set of three rest Hamiltonian constraints $T_1$ , $\theta $ and $\bar
\theta $. From the general algebraic viewpoint there are only two different
possibilities for the set of three arbitrary constraints: (I) all
constraints are of the first class, (II) one constraint is of the first
class and two are of the second class ones. The case under consideration
corresponds to the last situation for the free particle. Notice that the
coupling, being introduced into the constraints, may transform the
constraints of type (I) to those of type (II), whereas the inverse
transformation is highly impossible if a general configuration of external
fields is considered\footnote{%
There may exist, however, some special configurations of background fields for
which interaction (54) is consistent for another values of $\Delta
\;,\;m\;,\; \rho$. In particular, $\Delta$ may be arbitrary real
constant for
anti-de Sitter (de Sitter) space--time without torsion and electromagnetism
\cite{AdS} since the r.h.s. of Rel. (55) vanishes in the case.}. In other
words, the constraint structure of case (ii) is stable with respect to
general deformations by external fields.

Let us next consider the following generalization of the Hamiltonian
constraints (27--30): \addtocounter{equation}{1}
$$
T_1^{int}=\Pi^2 +\frac{g}{2} (e{\cal F}_{ab} {\cal M}^{ab} + \frac{1}{2}
{\cal R}_{abcd} {\cal M}^{ab} {\cal M}^{cd} + \Pi_c {\cal T}^c_{ab} {\cal M}%
^{ab} ) \approx 0 \eqno{(\theequation.a)}
$$
$$
\theta^{int} = \bar{\lambda}_{\dot{\alpha}} \Pi^{\dot{\alpha}\alpha}
\pi_{\alpha} \approx 0 \;\;,\;\;\bar{\theta}^{int} = \bar{\pi}_{\dot{\alpha}%
} \Pi^{\dot{\alpha}\alpha} \lambda_{\alpha} \approx 0 \eqno{(\theequation.b)}%
$$
$$
d^{int}=d \approx 0 \;\;,\;\; \bar{d}^{int}= \bar{d} \approx 0
\eqno{(\theequation.c)}%
$$
where, apart from the minimal covariantization, the nonminimal term is added
to the first constraint, being proportional to an arbitrary constant $g$.

In order to clarify the physical meaning of the parameter $g$, consider the
equations of motion for the coupled model described by the constraints (56).
To this end we, first, introduce the Dirac bracket associated with the pair
of the second class constraints $\theta$ and $\bar{\theta}$ by the rule
\addtocounter{equation}{1}
$$
\{A,B\}_{DB} =\{A,B\} -\frac{1}{2isM^2} \left( \{A,\theta\}\{\bar{\theta}%
,B\} - \{A,\bar{\theta}\}\{\theta,B\} \right) \eqno{(\theequation.a)}
$$
where
$$
M^2= -\frac{1}{2is} \{\theta,\bar{\theta} \} \approx m^2 + \frac{1+g}{2} (e%
{\cal F}_{ab} {\cal M}^{ab} + \frac{1}{2} {\cal R}_{abcd} {\cal M}^{ab}
{\cal M}^{cd} + \Pi_d {\cal T}^d_{ab} {\cal M}^{ab} ) \eqno{(\theequation.b)}
$$
may be regarded as an `effective' mass of the particle. Here we choose the
parameter $\rho$ to be pure imaginary constant $\rho= is$ as it leads to the
dynamics being invariant under the space-time reversions. Then the dynamics
of the system is completely determined by the evolution of extended momenta $%
\Pi_a$, space--time coordinates $x^a$ and spin tensor ${\cal M}_{ab}$. Note
that the latter satisfies the standard conditions \addtocounter{equation}{1}
$$
\Pi_a {\cal M}^{ab} \approx 0\;\;,\;\; {\cal M}_{ab} {\cal M}^{ab} \approx
2s^2 \eqno{(\theequation.a)}
$$
which allows to express this tensor via the vector of spin $\tilde{W}^a$
$$
\tilde{W}^a \equiv \frac{1}{2 \sqrt{-\Pi^2}} \epsilon^{abcd} {\cal M}_{bc}
\Pi_d\;\;,\;\;{\cal M}_{ab}=\frac{1}{2 \sqrt{-\Pi^2}} \epsilon_{abcd} \tilde{%
W}^c \Pi^d \eqno{(\theequation.b)}
$$
With the use of Dirac brackets (57) evolution of a phase space function $A$
is given by the equation \addtocounter{equation}{1}
$$
\dot A =\lambda \{A,T^{int}_1\}_{DB} \eqno{(\theequation)}
$$
$\lambda$ being an arbitrary Lagrange multiplier corresponding to the
reparametrization invariance. For the sake of simplicity we shall restrict
our consideration to the case of pure electromagnetic field. Then the
corresponding equations read \addtocounter{equation}{1}
$$
(\lambda e)^{-1} \dot{{\cal M}}_{ab} =g {{\cal F}_{ a }}^c {\cal M}_{ cb } +
\Pi_{ a } K_{ b } -(a\leftrightarrow b) \eqno{(\theequation.a)}
$$
$$
-(\lambda e)^{-1} \dot{\Pi}_{a} =2{\cal F}_{ab} \Pi^b +\frac{g}{2}
\partial_a {\cal F}^{bc} {\cal M}_{bc} +e {\cal F}_{ab} K^b
\eqno{(\theequation.b)}
$$
$$
(\lambda)^{-1} \dot{x}^{a} =2\Pi^a +e K^a \eqno{(\theequation.b)}
$$
where
$$
K_a = \frac{1}{M^2} {\cal M}_{ab} \left\{(2-g) {\cal F}^{bc} \Pi_c + \frac{g%
}{2} \partial^b {\cal F}^{cd} {\cal M}_{cd} \right\} \eqno{(\theequation.d)}
$$
$$
M^2= m^2 + \frac{1+g}{2} e{\cal F}_{ab} {\cal M}^{ab} \eqno{(\theequation.e)}
$$
Let us briefly discuss the obtained equations. First of all, one may see
that for the case of weak (that justifies an approximation $M^2 \approx m^2$%
) and constant electromagnetic field the first equation reduces to the well
known Frenkel-Nyborg equation for spin \cite{FrenkelNyborg,BagrovBord}. In
this case, the introduced parameter $g$ has the sense of Land\'e factor. As
is seen there are three special values of $g$ when the Eqs. (60) are
considerably simplified, namely $g = 0, 2, -1$. The first case corresponds
to the minimal coupling. The second one leads to the equations which, in the
case of constant electromagnetic field, describe the conventional precession
of spin tensor and four-velocity $\dot{x}^{a} =2\lambda\Pi^a$. Finally,
putting $g=-1$ one finds that $M^2 = m^2 $ that guarantees the
non-singularity of the Dirac bracket (57) on the whole phase space of the
system regardless of background field configurations.

Being associated with the constraints (56), the first-order (Hamiltonian)
action of the model looks like

\begin{equation}
S_H=\int \mbox{d}\tau (p_m\dot x^m+\pi _\alpha \dot \lambda ^\alpha +\bar
\pi _{\dot \alpha }{\bar \lambda }^{\dot \alpha }-\lambda T_1^{int}-\mu
\theta ^{int}-\bar \mu \bar \theta ^{int}-\nu d-\bar \nu \bar d)\,,
\end{equation}
where $\lambda ,\mu ,\bar \mu ,\nu ,\bar \nu $ are Lagrange multipliers. The
action can be readily brought to a second-order form by eliminating the
momenta $p_a,\pi _\alpha ,\bar \pi _{\dot \alpha }$ and Lagrange multipliers
with the aid of their equations of motion
$$
\frac{\delta S}{\delta p_a}=\frac{\delta S}{\delta \pi _\alpha }=\frac{%
\delta S}{\delta \bar \pi _{\dot \alpha }}=\frac{\delta S}{\delta \mu }=%
\frac{\delta S}{\delta \bar \mu }=\frac{\delta S}{\delta {\nu }}=\frac{%
\delta S}{\delta \bar \nu }=0
$$
It turns out, however, that in a general case this procedure leads to an
action functional with a rather complicated dependence upon the curvature
and torsion. Whereas, restricting to the case of minimal coupling in the
gravitational sector (i.e. neglecting the terms containing the Riemann
curvature and torsion in the constraints $T_1^{int}$), we come to the neat
generalization of the free model action \addtocounter{equation}{1}
$$
S={\displaystyle \int }d\tau {\cal L}\eqno{(\theequation.a)}
$$
$$
\begin{array}{c}
{\cal L}={\displaystyle \frac 1{4\lambda }}g_{mn}\dot x^m\dot x^n-\lambda
\left\{ m^2-4|\rho |^2{\displaystyle \frac{|\stackrel{\bullet }{\lambda }%
\!_\alpha {\lambda }^\alpha |^2}{(\dot x^me_{m\beta \dot \beta }\lambda
^\beta \bar \lambda ^{\dot \beta })^2}}\right\} + \\  \\
+\rho
{\displaystyle
{}{\frac{\dot x^me_{m\alpha \dot \alpha }\stackrel{\bullet
}{\lambda }\!_{}^\alpha \bar \lambda ^{\dot \alpha }}{\dot x^me_{m\beta \dot
\beta }\lambda ^\beta \bar \lambda ^{\dot \beta }}}^{}+\bar \rho {%
\displaystyle \frac{\dot x^me_{m\alpha \dot \alpha }{\lambda }^\alpha {%
\stackrel{\bullet }{\bar \lambda }}\!^{\dot \alpha }}{\dot x^me_{m\beta \dot
\beta }\lambda ^\beta \bar \lambda ^{\dot \beta }}}} \\  \\
-e{\cal A}_m\dot x^m
\end{array}
\eqno{(\theequation.b)}
$$
where
$$
\stackrel{\bullet }{\lambda }\!^\alpha =\dot \lambda ^\alpha -\frac
12(\lambda ge{\cal F}_{ab}+\dot x^m{\omega _m}_{ab})(\sigma ^{ab}{)_\beta }%
^\alpha \lambda ^\beta \eqno{(\theequation.c)}
$$
is the extended Lorentz-covariant derivative. Further, the case $g=0$ allows
to exclude explicitly the Lagrange multiplier $\lambda $ from the Lagrangian
by means of the equation $\delta S/\delta \lambda =0$:
\addtocounter{equation}{1}
$$
\begin{array}{c}
{\cal L}=\sqrt{-\dot x^2\left( m^2c^2-4|\rho |^2{\displaystyle \frac{|%
\stackrel{\bullet }{\lambda }\!_\alpha {\lambda }^\alpha |^2}{(\dot
x^me_{m\beta \dot \beta }\lambda ^\beta \bar \lambda ^{\dot \beta })^2}}%
\right) }+ \\  \\
+\;\rho
{\displaystyle {}{\frac{\dot x^me_{m\alpha \dot \alpha }\stackrel{\bullet
}{\lambda }{}^\alpha \!{}\,\bar \lambda ^{\dot \alpha }}{\dot x^me_{m\beta \dot \beta
}\lambda ^\beta \bar \lambda ^{\dot \beta }}}^{}+\bar \rho {\displaystyle
\frac{\dot x^me_{m\alpha \dot \alpha }{\lambda }^\alpha \stackrel{\bullet }{%
\bar \lambda }\!^{\dot \alpha }}{\dot x^me_{m\beta \dot \beta }\lambda ^\beta \bar \lambda^{\dot \beta }}}}- \\  \\
-e{\cal A}_m\dot x^m
\end{array}
\eqno{(\theequation)}
$$
The Lagrangian (63), along with the one for the free model (16) possesses
(for $\Delta =0$), besides $\lambda ^\alpha $--rescalings, the only gauge
symmetry -- reparametrization invariance.

\section{Quantization}

In this section we consider the operatorial formulation for the quantum
theory of the free model. We show that, for every case (i--i$\vee$), the
Hilbert space of the physical states is identified with the space of
corresponding unitary irreducible representation of the Poincar\'e group.

An explicit geometric construction underlying the model makes it possible to
operate in a manifestly covariant fashion, namely, to embed the Hilbert
space of physical states of the system into a space of smooth (spin)tensor
fields on ${\cal M}^6$. A complementary advantage of this approach is in its
direct relationship with the conventional on-shell realization for unitary
irreps of Poincar\'e group in terms of (spin)tensor fields on Minkowski
space.

We will be interested in those tensor fields on ${\cal M}^6$ which are
scalars in Minkowski space and tensor fields of type $\{k/2,l/2\}$ on the
sphere. This is equivalent to the following transformation property under
the action of Poincar\'e group (2.a),(6):
\begin{equation}
\Phi^{\prime}(x^{\prime},z^{\prime},{\bar z}^{\prime}) = \left(\frac{%
\partial z^{\prime}}{\partial z}\right)^{k/2} \left(\frac{\partial{\bar z}%
^{\prime}}{\partial{\bar z}}\right)^{l/2} \Phi(x,z,{\bar z}).
\end{equation}

The fields of this type will naturally arise if one starts with the particle
wave functions $\Psi(x^a,\lambda^{\alpha},\bar{\lambda}^{\dot{\alpha}})$
defined over the original configuration space ${\bf R}^{3,1}\times{\bf C}^2$
as scalar fields subject to the quantum kinematical constraints $\hat{d}$
and $\hat{\bar{d}}$: \addtocounter{equation}{1}
$$
(i\lambda^{\alpha} \frac{\partial}{\partial \lambda^{\alpha}} +\rho)
\Psi(x^a,\lambda^{\alpha},\bar{\lambda}^{\dot{\alpha}}) =0
\eqno{(\theequation.a)}
$$
$$
(i\bar{\lambda}^{\dot{\alpha}} \frac{\partial}{\partial \bar{\lambda}^{\dot{%
\alpha}}} +\bar{\rho}) \Psi(x^a,\lambda^{\alpha},\bar{\lambda}^{\dot{\alpha}%
}) =0. \eqno{(\theequation.b})%
$$
These equations imply that $\Psi$ is a homogeneous function in $%
\lambda^{\alpha},\bar{\lambda}^{\dot{\alpha}}$ and locally it can be
rewritten in terms of $\Phi^{\{i\rho/2,i\bar{\rho}/2\}}$ as
\addtocounter{equation}{1}
$$
\Psi(x^a,\lambda^{\alpha},\bar{\lambda}^{\dot{\alpha}}) = \kappa^{i\rho}
\bar{\kappa}^{i\bar{\rho}} \Phi^{\{i\rho,i\bar{\rho}\}} (x^a,z,\bar{z})
\eqno{(\theequation)}
$$
where the representation (21.a) for $\lambda^{\alpha}$ is used. Then the
transformation property (64) follows at once from (21.a) , (23) , (66).

It should be stressed that all the subsequent results could be obtained
without use of local coordinates $z$ by working directly in terms of
homogeneous functions over ${\bf R}^{3,1}\times {\bf C}^2$. However, the
tensor fields on ${\cal M}^6$ are convenient for our purpose due to their
transparent structure and well-realized decomposition with respect to the
Poincar\'e group \cite{ms particle}. Let us recall relevant facts conserning
the stracture of these fields.

The infinitesimal form of (64) is given by the generators
\addtocounter{equation}{1}
$$
{\bf P}_a = -{\rm i} \partial_a, \qquad {\bf J}_{ab} = -{\rm i}%
(x_a\partial_b - x_b\partial_a + M_{ab}), \eqno{(\theequation.a)}%
$$
where the spinning part of ${\bf J}_{ab}$ is realized by spherical variables
in the manner
$$
M_{ab} = (\sigma_{ab})_{\alpha\beta}M^{\alpha\beta} - ({\tilde\sigma}_{ab})_{%
{\dot\alpha}{\dot\beta}} \vspace{-12pt}{\bar M}^{{\dot\alpha}{\dot\beta}},
\eqno{(\theequation.b)}%
$$
$$
{}%
$$
$$
M^{\alpha\beta} = -z^\alpha z^\beta\partial_z + \frac{k}{2} \partial_z
(z^\alpha z^\beta), \qquad {\bar M}^{{\dot\alpha}{\dot\beta}} = - {\bar z}%
^{\dot\alpha} {\bar z}^{\dot\beta}\partial_{\bar z} + \frac{l}{2}
\partial_{\bar z}({\bar z}^{\dot\alpha} {\bar z}^{\dot\beta}).
\eqno{(\theequation.c)}%
$$

\noindent Then, two Casimir operators : the squared mass and the squared
spin $C^{\{k/2,l/2\}} = {\bf W}^a {\bf W}_a$, ${\bf W}^a = \frac{1}{2}%
\varepsilon^{abcd}{\bf P}_b {\bf J}_{cd}$ being the Pauli---Lubanski vector,
have the form \addtocounter{equation}{1}
$$
{\bf P}^a{\bf P}_a =-\Box \eqno{(\theequation.a)}%
$$
$$
C^{\{k/2,l/2\}} = -({\bf P},\xi)^2\partial_z\partial_{\bar z} + k({\bf P}%
,\xi)({\bf P},\partial_z\xi)\partial_{\bar z} + l({\bf P},\xi)({\bf P}%
,\partial_{\bar z}\xi)\partial_z -
$$
$$
- kl({\bf P}, \partial_z \xi)({\bf P},\partial_{\bar z}\xi) - {\bf P}^2
\left\{ \left(\frac{k-l}{2}\right)^2 + \frac{k+l}{2}\right\}.
\eqno{(\theequation.b)}%
$$

Let us consider the space $^{\uparrow }{\cal H}^{\{k/2,l/2\}}({\cal M}^6;m)$
of massive positive-frequency fields on ${\cal M}^6$ of the spherical type ${%
\{k/2,l/2\}}$ where $k$ , $l$ are integer. Such fields are defined to
satisfy the mass-shell condition \addtocounter{equation}{1}
$$
(\Box -m^2)\Phi ^{\{k/2,l/2\}}(x,z,\bar z)=0\eqno{(\theequation.a)}
$$
and possess the Fourier decomposition
$$
\Phi ^{\{k/2,l/2\}}(x,z,{\bar z})=\int \frac{d^3{\vec p}}{p^0}{\rm e}^{{\rm i%
}(p,x)}\Phi ^{\{k/2,l/2\}}(p,z,{\bar z}\vspace{-12pt}),%
\eqno{(\theequation.b)}
$$
where
$$
p^2+m^2=0,\qquad p^0>0.\eqno{(\theequation.c)}
$$
After passing to momentum representation with respect to $(x^a,p_b)$ the
second Casimir operator $C^{\{k/2,l/2\}}$ admits a simple geometric
interpretation. With every point of massive hyperboloid $p^2=-m^2$ one can
associate the smooth metric on $S^2$ \addtocounter{equation}{1}
$$
ds^2=\frac{4dzd{\bar z}}{(p^a\xi _a)^2}=2g_{z{\bar z}}dzd{\bar z},%
\eqno{(\theequation)}
$$
where $\xi _a$ is defined as in the Eq. (22). Since $p_a$ is a time-like
Lorentz vector, the Rel. (70) proves to define a smooth tensor field on $S^2$
of type \{-1, -1\}. This metric is covariant under the Lorentz
transformations (6), (9) in the following sense \addtocounter{equation}{1}
$$
{\frac{dz^{\prime }d{\bar z}^{\prime }}{(p_{\alpha \dot \alpha }^{\prime
}z^{\prime }{}^\alpha {\bar z}^{\prime }{}^{\dot \alpha })^2}}={\frac{dzd{%
\bar z}}{(p_{\alpha \dot \alpha }z^\alpha {\bar z}^{\dot \alpha })^2}},%
\eqno{(\theequation.a)}
$$
where
$$
{p^{\prime }}_{\alpha \dot \alpha }={N_\alpha }^\beta \left. {\overline{N}}%
_{\dot \alpha }\right. ^{\dot \beta }p_{\beta \dot \beta }%
\eqno{(\theequation.b)}
$$
and $z$ is transformed by the linear--fractional law related to the Lorentz
transformation by Rel.(9). The corresponding covariant derivatives $\hat
\nabla _z$ and $\hat \nabla _{\bar z}$ (acting on a tensor field of the
spherical type ${\{k/2,l/2\}}$) have the form \addtocounter{equation}{1}
$$
\hat \nabla _z=\partial _z+\frac k2\Gamma _{zz}^z\;\;,\;\;\Gamma
_{zz}^z=-2\partial _z\ln (p,\xi ),\eqno{(\theequation.a)}
$$
$$
\hat \nabla _{\bar z}=\partial _{\bar z}+\frac l2\Gamma _{\bar z\bar
z}^{\bar z}\;\;,\;\;\Gamma _{\bar z\bar z}^{\bar z}=-2\partial _{\bar z}\ln
(p,\xi ),\eqno{(\theequation.b)}
$$
and satisfy the commutation relation
\begin{equation}
[\hat \nabla _z,\hat \nabla _{\bar z}]=\frac 12(k-l)g_{z{\bar z}}R,\qquad
R=-p^2.
\end{equation}
The last expression manifests that, for every point of massive hyperboloid $%
p^2=-m^2$, the metric (70) is characterized by the constant positive
curvature $R=m^2$. Now, the squared spin operator (68.b) can be rewritten as
\addtocounter{equation}{1}
$$
C^{\{k/2,l/2\}}\equiv \Delta ^{\{k/2,l/2\}}=-2g^{z{\bar z}}\hat \nabla
_z\hat \nabla _{\bar z}+\frac{(k-l)}2\left( \frac{k-l}2+1\right) R.%
\eqno{(\theequation)}
$$
In this form the Casimir operator~ $C^{\{k/2,l/2\}}$ ~ is identified with a
special spherical Laplacian.

Employing the metric (70), the space $^{\uparrow }{\cal H}^{\{k/2,l/2\}}(%
{\cal M}^6;m)$ can be endowed with a Hilbert space structure where the
corresponding inner product is introduced as \addtocounter{equation}{1}
$$
\begin{array}{c}
< \Phi _1|\Phi _2 > _{\{k/2,l/2\}}= \\
\end{array}
$$
$$
=N\int \frac{d^3{\vec p}}{p^0}~\frac{dzd{\bar z}}{(p,\xi )^2}~\frac 1{(p,\xi
)^{k+l}}{\overline{\Phi _1^{\{k/2,l/2\}}(p,z,{\bar z})}}\Phi
_2^{\{k/2,l/2\}}(p,z,{\bar z}),\eqno{(\theequation)}
$$
with $N$ being some normalization constant.

It should be stressed that the inner product (75) is well defined on $S^2$
and is seen to be Poincar\'e invariant. As a result, the representation of
Poincar\'e group in the space $^{\uparrow }{\cal H}^{\{k/2,l/2\}}({\cal M}%
^6;m)$ is unitary. This representation can be readily decomposed into a
direct sum of irreducible ones by accounting already mentioned fact that the
spin operator $C^{\{k/2,l/2\}}$ coincides with the Laplacian $\Delta
^{\{k/2,l/2\}}$ which proves to be the Hermitian operator with respect to
the inner product (75). The spectrum of the Laplasian $\Delta ^{\{k/2,l/2\}}$
is given by the eigenvalues
\begin{equation}
s(s+1)R,\qquad s=\frac{|k-l|}2,\frac{|k-l|}2+1,\frac{|k-l|}2+2,\ldots
\end{equation}
The dimension of the eigenspace corresponding to an eigenvalue $s(s+1)R$
equals $2s+1$. This leads to the following decomposition
\begin{equation}
^{\uparrow }{\cal H}^{\{k/2,l/2\}}({\cal M}^6;m)={\mathop{\bigoplus}\limits%
_{s=\frac{|k-l|}2,\frac{|k-l|}2+1,\ldots }}~^{\uparrow }{\cal H}%
_s^{\{k/2,l/2\}}({\cal M}^6;m)
\end{equation}
Here the invariant subspace $^{\uparrow }{\cal H}_s^{\{k/2,l/2\}}({\cal M}%
^6;m)$ realizes unitary Poincar\'e representation of mass $m$ and spin $s$.
The expansion of an arbitrary field from $^{\uparrow }{\cal H}^{\{k/2,l/2\}}(%
{\cal M}^6;m)$, which corresponds to the decomposition (77), reads as
follows \addtocounter{equation}{1}
$$
\Phi ^{\{k/2,l/2\}}(p,z,{\bar z})=
$$
$$
=\sum_{s=\frac{|k-l|}2,\frac{|k-l|}2+1,\ldots }F_{\alpha _1\ldots \alpha
_{s-(k-l)/2}{\dot \alpha }_1\ldots {\dot \alpha }_{s+(k-l)/2}}(p)\times
\eqno{(\theequation.a)}
$$
$$
\times \frac{z^{\alpha _1}\ldots z^{\alpha _{s-(k-l)/2}}{\bar z}^{{\dot
\alpha }_1}\ldots {\bar z}^{{\dot \alpha }_{s+(k-l)/2}}}{(p,\xi )^{s-(k+l)/2}%
}
$$
\begin{sloppypar}

\noindent
In this expansion, each coefficient $F_{\alpha(s-(k-l)/2)\;{\dot\alpha}(s+
(k-l)/2)}(p)$ is symmetric in its undotted and, independently, in its
dotted indices
$$
F_{\alpha_1\ldots\alpha_{s-(k-l)/2}\,{\dot\alpha}_1\ldots
{\dot\alpha}_{s+(k-l)/2}}(p) = F_{\alpha_1\ldots\alpha_{s-(k-l)/2}\,
{\dot\alpha}_1\ldots {\dot\alpha}_{s+(k-l)/2}}(p),
\eqno{(\theequation.b)}$$
and $p$-transversal,
$$
p^{\dot{\beta}\beta} F_{\beta\alpha(s-(k-l)/2-1)\,{\dot\beta}{\dot\alpha}(s+
(k-l)/2-1)}(p) = 0.
\eqno{(\theequation.c)}$$
Together with the mass-shell condition $p^2=-m^2$ Eqs. (78.b,c) constitute the
set of relativistic wave equations for a massive field of spin $s$.
Thus, the $F$'s are identified with the Fourier transformations of
Poincar\'e-irreducible tensor fields on Minkowski space.
\end{sloppypar}

Now since we have recalled the necessary features of relativistic harmonic
analysis on ${\cal M}^6$, we are in a position to define the correct
operatorial formulation for the quantum theory of the model.

In Sec. 3, it has been established that all the physical observables of the
model, being the gauge invariant values, are the functions of the Poincar\'e
generators only. After quantization, every physical observable must be
assigned with a Hermitian operator acting in a Hilbert space. From this
point of view the quantization procedure for this model is equivalent to the
construction of unitary irreducible representation of Poincar\'e group with
the mass and spin (helicity for $m=0$) fixed by the constraints (33), (35).
As we have seen, for $m\neq 0$ such representations are naturally realized
in the spaces of massive positive frequency fields $^{\uparrow }{\cal H}%
^{\{k/2,l/2\}}({\cal M}^6;m)$. It seems instructive, however, to consider
this problem from the standpoint of the conventional Dirac quantization.

In the ordinary coordinate representation, the quantization of original
phase space variables is performed by the standard substitution
\begin{equation}
p_a\mapsto -i\partial _a\;\;,\;\;p_z\mapsto -i\partial _z\;\;,\;\;p_{\bar
z}\mapsto -i\partial _{\bar z}
\end{equation}
Then, the operators corresponding to Poincar\'e generators can be obtained
by the substitution of (79) into Hamiltonian generators (32.a),(41). Notice,
however, that there exists some inherent ambiguity in the ordering of $\hat
p_z$ , $\hat z$ and $\hat p_{\bar z}$ , $\bar z$ \thinspace due to the
nonlinear character of Lorentz transformations. Fortunately, the different
operator orderings in (41) result only in renormalization of the parameter $%
\rho $. So, in general, the quantum Poincar\'e generators has the form (67)
wherein $k$ and $l$ are arbitrary complex numbers. In what follows, however,
one has to restrict the parameters $k$ and $l$ to be integer, as for only
this case the infinitesimal Lorentz transformations are globally integrable
in the space of smooth tensor on ${\cal M}^6$ what is of the primary
importance for existence of a Hilbert space structure.

Thus, the Hilbert space of the system is embedded into the space ${\cal H}%
^{\{k/2,l/2\}}$ of smooth tensor fields of spherical type $\{k/2,l/2\}$.

Consider now the constraints (39). Once the space ${\cal H}^{\{k/2,l/2\}}$
has been chosen, we must require the quantum analogues of Hamiltonian
constraints to be well-defined operators on ${\cal H}^{\{k/2,l/2\}}$. Up to
a constant, this leads to the unique expression for quantum constraints.
Namely, the first class constraints $\hat{T}_1$ , $\hat{T}_2$ are naturally
associated with the Casimir operators for the Poincar\'e generators (67)
\addtocounter{equation}{1}
$$
\hat{T}_1 = p^2 + m^2 \eqno{(\theequation.a)}
$$
$$
\hat{T}_2 =-2g^{z\bar{z}} \hat{\nabla}_z \hat{\nabla}_{\bar{z}} -(\Delta^2 +
m^2(\Im m \rho)^2) -\alpha m^2 \eqno{(\theequation.b)}
$$
while the constraints $\hat{\theta}$ and $\hat{\bar{\theta}}$ can be
expressed in terms of covariant derivatives and metric
\begin{equation}
\hat{\theta} =-i\sqrt{2g^{z\bar{z}}} \hat{\nabla}_z \;\;\;\;,\;\;\;\; \hat{
\bar{\theta}} = -i\sqrt{2g^{z\bar{z}}} \hat{\nabla}_{\bar{z}}
\end{equation}
(here we have passed to the momentum representation with respect to $x^a$
and $p_a$). The arbitrary real constant $\alpha$ is introduced to account
the ambiguity in the ordering of covariant derivatives $\hat{\nabla}_z$ and $
\hat{\nabla}_{\bar{z}}$ (as is seen from (73)), that can be thought about as
a quantum correction to the constraint condition (39.b).

Consider now the quantization of all the special cases (i-i$\vee$) in turn.

\vspace{4mm}

\centerline{ Case (i)} \noindent There are two first class constraints $\hat
T_1$ and $\hat T_2$. Hence, the physical subspace is extracted by the
conditions \addtocounter{equation}{1}
$$
(p^2+m^2)\Phi _{phys}^{\{k/2,l/2\}}(p,z,\bar z)=0\eqno{(\theequation.a)}
$$
$$
(C^{\{k/2,l/2\}}-\Delta ^2-m^2(\Im m\rho ^2+\alpha ))\Phi
_{phys}^{\{k/2,l/2\}}(p,z,\bar z)=0\eqno{(\theequation.b)}
$$
Upon fulfillment of the mass-shell condition (82.a) the eigenvalues of
operators $C^{\{k/2,l/2\}}$ are given by (76). Hence, the equations (82)
admit a nontrivial solution only provided \addtocounter{equation}{1}
$$
\left( \frac \Delta m\right) ^2+(\Im m\rho )^2+\alpha =s(s+1)
$$
$$
s=\frac{|k-l|}2\;\;,\;\;\frac{|k-l|}2+1\;\;,\;\;\frac{|k-l|}2+2\ldots
\eqno{(\theequation.a)}
$$
Comparing (83.a) with the classical value for the squared spin (34) we get
the final relations
$$
\left( \frac \Delta m\right) ^2+(\Im m\rho )^2=s^2\;\;,\;\;\alpha =s%
\eqno{(\theequation.b)}
$$
Then, the positive-frequency solutions (i.e. $p^0>0$ ) of Eqs. (82) form the
Hilbert space $^{\uparrow }{\cal H}_s^{\{k/2,l/2\}}({\cal M}^6;m)$.

\vspace{4mm} \centerline {\ Case (ii) } \noindent Besides the mass-shell
condition (80.a) there are two constraints (81) satisfying the algebra
\begin{equation}
[\hat \theta ,\hat{\bar \theta} ]=(k-l)p^2
\end{equation}
We restrict $k$ , $l$ to satisfy
\begin{equation}
|k-l|=|\Im m\rho |
\end{equation}
This condition implies that the classical and quantum spin values should
coincide and equal $|\Im m\rho |$, as we show below.

Thus, we have the set of one first (80.a) and two second class constraints
(81). A correct definition of physical states will be achieved if they are
required to be annihilated by the first-class constraint operator
\addtocounter{equation}{1}
$$
(p^2+m^2)\Phi _{phys}^{\{k/2,l/2\}}(p,z,\bar z)=0\eqno{(\theequation.a)}
$$
as well as by the half of the second class constraints
$$
\mbox {for}\;k\geq l,\;\;\; \hat{\bar \theta}
 \;\Phi _{phys}^{\{k/2,l/2\}}(p,z,\bar z)=0\eqno{(\theequation.b)}
$$
$$
\mbox {for}\;k\leq l,\;\;\;\hat \theta \;\Phi _{phys}^{\{k/2,l/2\}}(p,z,\bar
z)=0\eqno{(\theequation.c)}
$$
As a consequence of Eqs. (74, 86), the positive frequency solutions of Eqs.
(86) $^{}{\Phi }_{phys}^{\{k/2,l/2\}}$ satisfy
\begin{equation}
{\Phi }_{phys}^{\{k/2,l/2\}}\in \;{^{\uparrow }{\cal H}_s^{\{k/2,l/2\}}(%
{\cal M}^6;m)}\;\;,\;\;s=\frac 12|k-l|
\end{equation}
Reversely, each $\Phi ,\Phi \in {^{\uparrow }{\cal H}}_{|k-l|/2}^{\{k/2,l/2%
\}}({\cal M}^6;m)$ satisfies to Eqs. (86) as it is seen from the following
sequence (for a moment, we take $k>l$ for definiteness)
\addtocounter{equation}{1}
$$
0=<\Phi |C^{\{k/2,l/2\}}-m^2s(s+1)|\Phi >_{\{k/2,l/2\}}=<\Phi |\hat \theta
\; \hat{\bar \theta} |\Phi >_{\{k/2,l/2\}}=
$$
$$
=||\hat{\bar \theta} \Phi ||_{\{k/2,l/2\}}^2\Rightarrow \hat{\bar \theta}
 \Phi =0%
\eqno{(\theequation.a)}
$$
Thus the subspace of smooth positive-frequency solutions to Eqs. (86) is
identified with the space of unitary irreducible representation of the
Poincar\'e group $^{\uparrow }{\cal H}_{|k-l|/2}^{\{k/2,l/2\}}({\cal M}^6;m)$%
, being the lowest member of the direct sum (77).

\vspace{4mm} \centerline {Case (iii)} \noindent This case corresponds to the
massless particle with discrete helicity $\Im m\rho =0,\pm 1/2,\pm 1,$ $%
\!\pm 3/2,\pm 2,\ldots $ It is characterized by the set of three first-class
constraints $\hat T_1,\hat \theta $ and $\bar \theta $. The physical wave
functions are subject to the conditions \addtocounter{equation}{1}
$$
\hat{\bar \theta} \Phi _{phys}^{\{k/2,l/2\}}(p,z,\bar z)=0\eqno{(\theequation.a)}
$$
$$
\hat \theta \Phi _{phys}^{\{k/2,l/2\}}(p,z,\bar z)=0\eqno{(\theequation.b)}
$$
$$
p^2\;\Phi _{phys}^{\{k/2,l/2\}}(p,z,\bar z)=0\eqno{(\theequation.c)}
$$
Note that, for $k\neq l$, the constraint (89.c) arises as a consistency
condition for the Eqs. (89.a,b) by virtue of (84). The important fact is
that Eqs. (89.a-c) do not possess smooth solutions for arbitrary $k$ , $l$,
so one has to define correctly the functional space where the solution is
sought. To formulate the right conditions let us introduce the mapping $\hat
\pi $ by the rule \addtocounter{equation}{1}
$$
\mbox {for}\;k\geq l,\;\;\;\hat \pi :\Phi _{phys}^{\{k/2,l/2\}}\to \Phi
_{phys}^{\{(k-l)/2,0\}}=(p,\xi )^{-l}\Phi _{phys}^{\{k/2,l/2\}}%
\eqno{(\theequation.a)}
$$
$$
\mbox {for}\;k\leq l,\;\;\;\hat \pi :\Phi _{phys}^{\{k/2,l/2\}}\to \Phi
_{phys}^{\{0,(l-k)/2\}}=(p,\xi )^{-k}\Phi _{phys}^{\{k/2,l/2\}}%
\eqno{(\theequation.b)}
$$
Then the right restriction on the functional space is the requirement for $%
\hat \pi \Phi _{phys}^{\{k/2,l/2\}}$ to be a smooth tensor field on $S^2$.
Choose, for a moment, $k\geq l$. Under the mapping $\hat \pi $ any solution
of Eqs. (89.a-c) transforms to that for

\addtocounter{equation}{1}
$$
(p,\xi) \partial_{\bar{z}} (\hat{\pi} \Phi)^{\{(k-l)/2,0\}}= 0
\eqno{(\theequation.a)}%
$$
$$
\hat{\theta} (\hat{\pi} \Phi)^{\{(k-l)/2,0\}}_{phys} = \left((p,\xi)
\partial_{z}-(k-l)(p,\partial_z \xi)\right) (\hat{\pi} \Phi)^{\{(k-l)/2,0%
\}}_{phys}= 0 \eqno{(\theequation.b)}%
$$
According to the Riemann-Roch theorem \cite{RR theorem}, the space of smooth
solutions for Eq. (91.a) (for every $p_a$) is $(k-l+1)$-dimensional space of
holomorphic tensor fields of the form \addtocounter{equation}{1}
$$
\Phi^{((k-l)/2,0)} (p,z)= \Phi_{ \alpha_1\ldots\alpha_{k-l} } (p)
z^{\alpha_1}\ldots z^{{\alpha}_{k-l}} \eqno{(\theequation)}%
$$
Substituting (92) into (91.b) and making use of the identity
\begin{equation}
z^{\alpha} \partial_z z^{\beta} - z^{\beta} \partial_z z^{\alpha} =\epsilon
^{\alpha\beta}
\end{equation}
one gets \addtocounter{equation}{1}
$$
p^{\dot{\beta} \alpha_1 } \Phi_{\alpha_1 \ldots \alpha_{k-l}} (p) =0
\eqno{(\theequation)}%
$$
Here one immediately recognizes the relativistic wave equation for a
massless field of helicity $(k-l)/2$.

The treatment of the case $k\leq l$, being analogous to the above, gives the
result \addtocounter{equation}{1}
$$
\Phi^{\{0,(l-k)/2\}} (p,z)= \Phi_{\dot{\alpha}_1\ldots \dot{\alpha}_{l-k}}
(p) \bar{z}^{\dot{\alpha}_1}\ldots \bar{z}^{\dot{\alpha}_{l-k}}
\eqno{(\theequation.a)}%
$$
$$
p^{\dot{\alpha}_1 \beta } \Phi_{\dot{\alpha}_1 \ldots \dot{\alpha}_{l-k}} (p)
=0 \eqno{(\theequation.b)}%
$$
presenting wave functions of helicity $-1/2(l-k)$ field.

Comparing the above formulas (94),(95.b) with the classical expression for
helicity (35) we set
\begin{equation}
\frac{k-l}2=\Im m\rho
\end{equation}
Furthermore, one observes that a general solution to Eqs. (89.c),(94),(95.b)
has the form \addtocounter{equation}{1}
$$
\Phi _{\alpha _1\ldots \alpha _{k-l}}(p)=\eta _{\alpha _1}\ldots \eta
_{\alpha _{k-l}}\Phi (p)\eqno{(\theequation.a)}
$$
$$
\Phi _{\dot \alpha _1\ldots \dot \alpha _{k-l}}(p)=\bar \eta _{\dot \alpha
_1}\ldots \bar \eta _{\dot \alpha _{l-k}}\bar \Phi (p)\eqno{(\theequation.b)}
$$
where
$$
p_{\alpha \dot \alpha }=\eta _\alpha \bar \eta _{\dot \alpha }%
\eqno{(\theequation.c)}
$$
is the standard twistor representation of a light-like four vector $%
p_{\alpha \dot \alpha }$ through a commuting Weyl spinor $\eta _\alpha $
determined by the Eq. (97.c) modulo a phase: $\eta _\alpha \sim e^{i\chi
}\eta _\alpha $. Accounting Eqs. (90),(92),(95),(97) as well as the equality
$(p,\xi )=(\eta _\alpha z^\alpha )(\bar \eta _{\dot \alpha }\bar z^{\dot
\alpha })$ one may come to the following general solution to the original
Eqs. (89) as
\begin{equation}
\Phi _{phys}^{\{k/2,l/2\}}(p,z,\bar z)=(\eta _\alpha z^\alpha )^k(\bar \eta
_{\dot \alpha }\bar z^{\dot \alpha })^l\Phi (p)\;\;,\;\;p^2=0
\end{equation}
Then the relevant realization for the norm, providing positive-frequency
wave functions (98) with a Hilbert space structure, is simply built as
\begin{equation}
< \Phi | \Phi > =N\int \frac{d^3{\vec p}}{p^0}~{\overline{\Phi }\Phi }%
\;\;,\;\;p^0=\sqrt{\vec p^2}
\end{equation}
where $N$ is a normalization constant. Note that the phase ambiguity for $%
\Phi $ arising from that for $\eta _\alpha $ does not contribute to this
norm, as it should be.

\vspace{4mm} \centerline {Case (i$\vee $)} \noindent This time the model
describes a particle with continuous helicity $\Delta \;,\;0<\Delta <\infty $%
. The dynamics is characterized by two abelian first-class constraints
(39.a,b). The quantum constraints $\hat T_1\;,\;\hat T_2$ and the physical
states $\Phi _{phys}^{\{k/2,l/2\}}$ are defined for arbitrary $k$ and $l$ as
in the case (i) provided that $m=0$ : \addtocounter{equation}{1}
$$
p^2\;\Phi _{phys}^{\{k/2,l/2\}}(p,z,\bar z)=0\eqno{(\theequation.a)}
$$
$$
(2g^{z\bar z}\hat \nabla _z\hat \nabla _{\bar z}+\Delta ^2)\;\Phi
_{phys}^{\{k/2,l/2\}}(p,z,\bar z)=0\eqno{(\theequation.b)}
$$
Note that, as a consequence of Eq. (73),
\begin{equation}
[\hat \nabla _z,\hat \nabla _{\bar z}]=0
\end{equation}
Hence, the ordering ambiguity does not appear for the constraint $\hat T_2$.
For the purposes of quantization it is enough to deal with the even $k-l$.
Then, the operators $i\Delta \hat \theta =i\Delta (p,\xi )\hat \nabla _z$
and $i\Delta \hat{\bar \theta} =i\Delta (p,\xi )\hat \nabla _{\bar z}$ are
inverse to each other on the space of solutions to Eqs. (91), that makes
possible to define the smooth one-to-one mapping \addtocounter{equation}{1}
$$
\Phi _{phys}^{\{k/2,l/2\}}\to \Phi
_{phys}^{\{(k+l)/2\,,\,(k+l)/2\}}=(i\Delta \hat \theta )^{(k-l)/2}\Phi
_{phys}^{\{k/2,l/2\}}\eqno{(\theequation.a)}
$$
$$
\Phi _{phys}^{\{(k+l)/2\,,\,(k+l)/2\}}\to \Phi
_{phys}^{\{k/2,l/2\}}=(i\Delta \hat{\bar \theta} )^{(k-l)/2}\Phi
_{phys}^{\{(k+l)/2\,,\,(k+l)/2\}}\eqno{(\theequation.b)}
$$
Furthermore, for $k=l$ the physical subspaces are identified with $\Phi
_{phys}^{\{0,0\}}$ by means of the formula \addtocounter{equation}{1}
$$
\Phi _{phys}^{\{k/2,k/2\}}\to \Phi _{phys}^{\{0,0\}}=(p,\xi )^{-k}\Phi
_{phys}^{\{k/2,k/2\}}\eqno{(\theequation)}
$$
The last mapping is smooth except the point $(p,\xi )=0$ \footnote{%
In the distinction to the massive case the spherical metric $g_{z\bar{z}} =
2(p,\xi)^{-2}$ is no longer smooth but has one singular point when $p_a \sim
\xi_a$. From Eq. (73) it follows that the curvature of the metric $g_{z\bar{z%
}}$ equals zero. In the special reference system $p^a = (p^0 ,0,0, p^0)$ one
finds the euclidean plane metric $g_{z\bar{z}} =\frac{1}{2}p_0^{-2}$ and the
corresponding singular point on $S^2$ is identified with the infinitely
removed point $z=\infty$.} and we restrict the functional space for
arbitrary $k$ , $l$ by requiring the image of two consequent mappings
(102.a), $(103)$ to be a smooth function on $S^2$. For $\Phi
_{phys}^{\{0,0\}}\equiv \Phi $, the Eqs. (100) take the form
\begin{equation}
((p,\xi )^2\partial _z\partial _{\bar z}+\Delta ^2)\Phi =0\;\;,\;\;p^2=0
\end{equation}
The smooth positive-frequency solutions to the Eq. (104) form a Hilbert
space with respect to the measure (75) (wherein $k=l=0$) taken at $m=0$
\footnote{%
It is also interesting to note that this case is characterized by the same
constraints and
possesses the same number of degrees of freedom as the massive one (i).
Thus, it could be treated as a massless limit of the massive
particle model (i).}:
\begin{equation}
< \Phi |\Phi > =N\int \frac{d^3{\vec p}}{p^0}~\frac{dzd{\bar z}}{(p,\xi
)^2}\overline{\Phi }\Phi \;\;,\;\;p^0=\sqrt{{\vec p}^2}
\end{equation}
Indeed, despite $(p,\xi )^{-2}$ has the singular point, one may rewrite
(105) using Eq. (94) as
\begin{equation}
< \Phi |\Phi > =-\Delta ^{-2}N\int \frac{d^3{\vec p}}{p^0}~dzd{\bar z}%
\overline{\Phi }\partial _z\partial _{\bar z}\Phi =\Delta ^{-2}N\int \frac{%
d^3{\vec p}}{p^0}~dzd{\bar z}|\partial _z\Phi |^2
\end{equation}
which is obviously well-defined since $\Phi $ is a smooth function on $S^2$.
Hence, the Poincar\'e group representation, being realized on the $\Phi
_{phys}^{\{0,0\}}$, is unitary. To prove the irreducibility, it is suitable
to apply the Wigner method of a little group and to make sure that the
representation of the momentum $p_a\;(p^2=0)$ stability subgroup (little
group) is irreducible in the subspace $|p_a>\equiv \Phi
_{phys}^{\{0,0\}}(p,z,\bar z)$. The little group of a light-like vector is
known to be isomorphic to the group $E(2)$ of motions of two-dimensional
euclidean plane, and the apparent proof of the fact is that for the special
$%
p_a$ of the form $p^a=(p^0,0,0,p^0)$ the metric becomes flat and the Eq.
(104) takes the form of the ordinary one determining eigenfunctions for the
Laplace operator on a plane \footnote{%
The stability subgroup of a momentum $p_a$ coincides with the invariance
group of the Eq. (104) at fixed $p_a$.}
\begin{equation}
(\partial _z\partial _{\bar z}+\Delta ^2)\Phi (\stackrel{\circ }{p}_a,z,\bar
z)=0
\end{equation}
The restriction of the norm (105) to the subspace $|\stackrel{\circ }{p}_a>$
yields the ordinary $L^2$-norm for the functions on a plane. The smooth
solutions to Eq.(101), being normalizable with respect to $L^2$-norm, prove
to form the $E_2$-irreducible infinite-dimensional Hilbert space and can be
written in terms of the first-kind Bessel functions $J_\nu \;,\;\nu
=0,1,2,\ldots ,$ \cite{Vilenkin} as follows \addtocounter{equation}{1}
$$
\Phi (\stackrel{\circ }{p}_a,z,\bar z)=\sum\limits_{\nu =0}^\infty c_\nu
(-i)^\nu \left( \frac z{\bar z}\right) ^{\displaystyle \frac \nu 2}J_\nu
(\Delta |z|)\eqno{(\theequation.a)}
$$
where $c_\nu $ are complex numbers obeying the restriction
$$
\sum\limits_{\nu =0}^\infty |c_\nu |^2\;<\infty \eqno{(\theequation.b)}
$$

This completes the proof that, in this case, the model describes the
relativistic massless particle with continuous helicity $\Delta $. Note,
that it is the same little group technique which reduces the proof of
irreducibility and unitarity to the analogous proof for the stability
subgroup representation, can be employed for all the cases (i),(ii),(iii),
but  we have preferred there to keep the manifest Poincar\'e--invariance for
the reasons of aesthetics. As to the explicit Poincar\'e--covariant
description for (i$\vee $), note, that, since the representation of the
little group is infinite-dimensional, the description of the physical wave
function in terms of Lorentz tensors is irrelevant and, thus, the approach
used above seems to be the most appropriate to the case.

\section{ Conclusion}

Let us briefly overview and comment the results. In this paper we have
considered the most general Poincar\'e invariant mechanical system (without
higher derivatives) with configuration space ${\cal M}^6$. We have shown,
that the requirement of relativistic mass-shell conditions to arise in the
theory, determines the reparametrization invariant action of the theory up
to the four arbitrary constant parameters. The model can be conceptually
treated as the universal model of relativistic spinning particle in $d=4$
Minkowski space.

Depending on a choice of the parameters, the model is characterized by a
certain number of Hamiltonian constraints providing the identical (on the
constraint surface) conservation of mass and spin (helicity) of the
particle. On the other hand, the number and the structure of the arising
constraints is precisely that which provides the correct number of physical
degrees of freedom for the corresponding particle.

The remarkable feature of the model is the existence (in the massive case)
of special points in the space of the model parameters where one of the
first class constraints disappears, being substituted by two second class
ones, that provides the different description of the same physical
situation, i.e. the massive spinning particle. The significance of this
special formulation becomes evident by the observation that there is the
constraint structure in the case which can be noncontradictory deformed, and
thereby the model allows to introduce a consistent interaction of the
particle to arbitrary external electromagnetic and gravity fields. It should
be mentioned that the proposed method leaves a wide freedom in the choice of
possible consistent interactions. We have considered in more details the
particular type of nonminimal interaction which leads to the description of
a massive spinning particle with arbitrary fixed giromagnetic ratio.

It is interesting to note that for some special values of giromagnetic ratio
$g=0,2,-1$ the interaction is symplified. The first case corresponds to the
minimal coupling. The second one leads to the equations which, in the case
of constant electromagnetic field, describe the conventional precession of
spin tensor and four-velocity $\dot x^a=2\lambda \Pi ^a$. Finally, putting $%
g=-1$ one finds that $M^2=m^2$ that guarantees the non-singularity of the
Dirac bracket (57) on the whole phase space of the system regardless of
background field configurations. It should be noted that these values are
special only from the standpoint of classical spinning particle dynamics and
tell us nothing about the values of giromagnetic ratio in (unknown yet)
consistent higher-spin field theories. In fact, a variety of interactions,
being admissible in the classical regime for the particle, corresponds to
the set of higher spin equations coupled to electromagnetic field. At the
mean time, the free Lagrangian massive higher spin field theories contain a
subtle auxiliary field structure \cite{higherspins}, that makes almost all
these naive interactions to be contradictory. Therefore, the question of
`the true' giromagnetic ratio value may probably have another answer from
the standpoint of the field theory.

We have thoroughly performed the covariant operatorial quantization of the
model and have explicitly constructed the corresponding Hilbert spaces which
are shown to coincide with the spaces of unitary irreducible representations
of the Poincar\'e group.

The constructed model reveals the properties of universal and minimal model
for $D=4$ spinning particle. The model is universal since it's configuration
space is mass- and spin- independent; it is minimal since its configuration
manifold ${\cal M}^6$ turns out to have the minimal possible dimension among
the homogeneous spaces of the Poincar\'e group admitting a nontrivial
dynamics for spin.

Perhaps, this universality can open a way to consider the joint dynamics of
the higher-spin fields both at the classical and quantum levels. The
promissing example of such a description is that the relativistic harmonic
analysis allows to treat a massive tensor field on ${\cal M}^6$ as a direct
sum of massive Poincar\'e-irreducible spintensor fields on Minkowski space.
If it is possible to find the analogous phenomena beyond the free level then
one could try to construct the consistent interaction between higher spins
in terms of tensor fields on ${\cal M}^6$.

The proposed model admits the generalization to arbitrary dimension. The
basic problem is to choose an appropriate inner space. For example, for $D=3$
the universal model of anyon has been derived with $S^1$ as the inner space
\cite{D3}. For $D=6$ the corresponding universal model could be constructed
with the inner space being complex projective space ${\bf C}P^3$ \cite{D6}.

The possibility to introduce a consistent interaction in the massless case
still remains in question within the proposed construction, since the
constraint structure does not survive nontrivial deformations when the mass
vanishes.

\section{Acknowledgments}

The authors would like to thank V.G. Bagrov for discussions and useful
references. We wish to express our special gratitude to S.M. Kuzenko for
significant advice and stimulating criticism.

The work is supported in part by European Community Commission (contract
INTAS-93-2058) and by International Soros Science Education Program
(grants No. d1312, a593-f, a634-f).

\end{document}